\documentclass{amsart}[12pt]

\usepackage{amsmath,amsfonts,latexsym,amssymb}
\usepackage{epsfig,epstopdf,color}
\newcommand{\rem}[1]{}

\definecolor{Violet}{cmyk}{0.79,0.88,0,0}
\definecolor{Plum}{cmyk}{0.50,1,0,0}
\definecolor{Periwinkle}{cmyk}{0.57,0.55,0,0}
\definecolor{ForestGreen}{cmyk}{0.91,0,0.88,0.12}
\definecolor{OliveGreen}{cmyk}{0.64,0,0.95,0.40}
\definecolor{BrickRed}{cmyk}{0,0.89,0.94,0.28}
\definecolor{DarkOrchid}{cmyk}{0.40,0.80,0.20,0}
\definecolor{Fuchsia}{cmyk}{0.47,0.91,0,0.08}
\definecolor{Mulberry}{cmyk}{0.34,0.90,0,0.02}
\definecolor{Maroon}{cmyk}{0,0.87,0.68,0.32}
\definecolor{Mahogany}{cmyk}{0,0.85,0.87,0.35}
\definecolor{RawSienna}{cmyk}{0,0.72,1,0.45}
\definecolor{YellowOrange}{cmyk}{0,0.42,1,0}
\definecolor{OrangeOrange}{cmyk}{0,0.62,1,0}
\definecolor{BurntOrange}{cmyk}{0,0.51,1,0}
\definecolor{Bittersweet}{cmyk}{0,0.75,1,0.24}
\definecolor{RedOrange}{cmyk}{0,0.77,0.87,0}
\definecolor{Sepia}{cmyk}{0,0.83,1,0.70}
\definecolor{Brown}{cmyk}{0,0.81,1,0.60}
\definecolor{Tan}{cmyk}{0.14,0.42,0.56,0}
\definecolor{darkorange}{cmyk}{.20,.50,.80,0}    
\definecolor{lightorange}{cmyk}{.07,.37,.65,0}   
\definecolor{darkpeagreen}{cmyk}{.50,.30,.50,0}  

\newtheorem{proposition}{Proposition}

\newcommand{\rhobar}{\overline{\rho}}
\newcommand{\kappabar}{\overline{\kappa}}
\newcommand{\bA}{\boldsymbol{A}}

\newcommand{\bq}{\boldsymbol{q}}

\newcommand{\bv}{\boldsymbol{v}}
\newcommand{\bw}{\boldsymbol{w}}
\newcommand{\bx}{\boldsymbol{x}}

\def\contract{\makebox[1.2em][c]{\mbox{\rule{.6em}
{.01truein}\rule{.01truein}{.6em}}}}
\newcommand{\bfi}{\bfseries\itshape}

\newcommand{\remfigure}[1]{#1}

\makeatletter
\@addtoreset{equation}{section}

\makeatother

\rem{
\pagestyle{myheadings}
\markright{18 Sept 2006 
\hfil Geometric order parameter equations}
}

\begin{document}

\title[Formation and evolution of singularities in anisotropic geometric continua]
{\LARGE
Formation and evolution of singularities in anisotropic geometric continua
}

\author{D. D. Holm}
\address{
(D. D. Holm)
Department of Mathematics, Imperial College London SW7 2AZ, UK
{email: d.holm@ic.ac.uk}
\linebreak
and
\linebreak
Computer and Computational Science, MS D413,
Los Alamos National Laboratory,
Los Alamos, NM 87545, USA
{email: dholm@lanl.gov}
}

\author{V. Putkaradze}
\address{
(V. Putkaradze)
Department of Mathematics, Colorado State University
{email: putkarad@math.colostate.edu}
}

\date{Started May 30, 2006, Today's version November 14, 2006}
\date{\today}

\maketitle

\begin{abstract} \LARGE
Evolutionary PDEs for geometric order parameters that admit propagating singular solutions are introduced and discussed. These singular solutions arise as a result of the competition between nonlinear and nonlocal processes in various familiar vector spaces. Several examples are given. The motivating example is the directed self assembly of a large number of particles for technological purposes such as nano-science processes, in which the particle interactions are anisotropic. This application leads to the derivation and analysis of gradient flow equations on Lie algebra valued densities. The Riemannian structure of these gradient flow equations is also discussed.

\end{abstract}
\LARGE

\tableofcontents

\section{Introduction} 
Many physical processes may be understood as aggregation of individual
`components' at a variety of scales into a final `product'. Diverse examples of
such processes  include the formation of stars, galaxies and solar
systems at large scales, organization of insects and organisms into colonies at
mesoscales and self-assembly of proteins, nanotubes or micro/nanodevices at
micro- and nanoscales  \cite{Whitesides2002}. Some of these processes,
such as  nano-scale self-assembly of molecules and particles are of great
technological interest \cite{Rabani2003,XB2004}. In most practical cases, the 
assembling pieces do not have spherical symmetry, and the interaction between those pieces is not \emph{central}, i.e., it is dependent not only on the density of the particles, but also on their mutual orientation. A recent example, motivating this paper,  is
 self-assembly of non-circular floating particles (squares, hexagons \emph{etc.})
\cite{Boden1999,Grzybowski2001,Grzybowski2004}.  Due to the large number of
particles involved in self-assembly for technological purposes, ($10^9-10^{12}$), the development of continuum descriptions for aggregation or self-assembly is a natural approach toward theoretical understanding and modeling. 
\\ 

Progress has been made recently in the derivation of continuum evolutionary equations for self-assembly of molecules evolving under a long-range central interaction potential. The resulting continuum models emerge in a class of partial differential equations (PDEs)  that are both nonlinear and nonlocal  \cite{HoPu2005,HoPu2006}. 
Nonlinearity and nolocality may combine in an evolutionary process to
form coherent singular structures that propagate and interact amongst
themselves. Thus, self-assembly is modeled in the continuum description as the emergent formation of singular solutions. Classic examples of equations admitting such singular solution behavior include the Poisson-Smoluchowsi (PS) equation \cite{PoSm-ref}, Debye-H\"uckel equations \cite{DeHu1923} and the Keller-Segel (KS) system of equations \cite{KeSe-ref}. PS and KS are gradient flows for densities that collapse to subspaces in finite time. Their respective collapses model the formation of clumps from randomly walking sticky particles (PS) and the development of concentrated biological patterns via chemotaxis (KS). For recent thorough review and references, see \cite{Ho2003}.  
A variant of these equations of potential use in modeling formation of
clustering in biology (e.g., formation of herds of antelope, or schools
of fish) was introduced in Bertozzi and Topaz \cite{BeTo2006}. Dynamics and collapse of self-gravitating systems has been studied by Chavanis, Rosier and Sire \cite{Chavanis2002}. Finally, aggregation of particles with the applications to directed self-assembly in nanoscience was considered by Holm and Putkaradze (HP) in
\cite{HoPu2005,HoPu2006}. 
The singular solutions of the latter HP
equation emerge as delta functions supported on a subspace from smooth initial conditions. These clumps or ``clumpons'' then propagate and interact by aggregating into larger (more massive) clumps when they collide.   See also \cite{GiTi2005} for a thoughtful discussion of singularity formation in fundamental mathematical terms. \\

Remarkably, the aggregation and emergence of singular solutions from smooth initial data due to nonlinearity and nonlocality need not be dissipative. A Hamiltonian example is provided by the Camassa-Holm (CH) equation for shallow water waves in its limit of zero linear dispersion \cite{CaHo1993}. In this limit, the CH equation describes geodesic motion on
the Lie group of smooth invertible maps with smooth inverses (the
diffeomorphisms, or diffeos, for short) with respect to the $H^1$
Sobolev metric. Being an integrable Hamiltonian system, the CH equation has soliton solutions which emerge in its initial value problem. In its dispersionless limit these CH soliton solutions develop a sharp peak at which the derivative is discontinuous, so the second derivative is singular. At the positions of these propagating peaks (the peakons),  the momentum density is concentrated into delta functions. The CH peakons propagate and interact elastically by exchanging momentum; so they bounce off each other when they collide. \\

CH arises as an Euler-Poincar\'e (EP) equation from Hamilton's principle defined using a Lagrangian which is right-invariant under the action of the diffeos on their own Lie algebra (the tangent space at the identity) \cite{HoMaRa1998}. In the EP approach, variations in Hamilton's principle of the vectors in linear representation spaces of the group of diffeos are induced from variations of the
diffeos themselves. Hence these variations in the EP Hamilton's
principle arise as geometrical operations, primarily as Lie derivatives of elements of the appropriate linear representation spaces. Moreover, evolution by CH -- or more generally by any of the class of EP equations on the diffeos (EPDiff) -- is by coadjoint action of the diffeos on the momentum density. This infinitesimal action is defined as the Lie derivative of momentum density by its corresponding velocity vector field. The momentum density for a singular solution of CH is supported on its singular set, regarded as an embedded submanifold \cite{HoMa2004}. \\

CH and other equations in the class of EPDiff equations hint that the actions of diffeos may be taken as a paradigm for the development and propagation of singularities in other geometrical quantities, or at least would provide a framework for studying the process and discovering additional examples of it. In fact, the HP equation was discovered by following Otto \cite{Otto2001} in formulating the process of nonlinear nonlocal evolution using a variational principle in which the variations were induced by the infinitesimal actions of diffeos on densities \cite{HoPu2005,HoPu2006}. Just as in Hamilton's principle for the EP equations, this infinitesimal action is the Lie derivative with respect to a vector field and for HP the vector field was related by Darcy's Law to the flux of density. \\

The derivation of evolution equations for density in the framework of Darcy's law is now well established: The velocity of a particle is taken to be proportional to the force acting on it, and the conservation law for density $\rho$ readily establishes a PDE for its evolution. The Darcy law evolution equation for density corresponds to conservation of the $n$-form in $n$-dimensional space $\rho \mbox{d}^n \mathbf{x}$ along characteristics of a velocity determined from the density and its gradient. Physically, the conservation law for $\rho \mbox{d}^n \mathbf{x}$ means preservation of the number of particles, or mass, in an infinitesimal volume. 
\\

The density evolution equation closes if the potential of interaction between the particles depends only on their relative position. This framework is simple and attractive. However, the energy of some physical systems depends strongly on additional geometric quantities, such as the mutual orientations of their \emph{pieces}. Examples of such systems are numerous and range from micro-biological applications (mutual attraction of cells, viruses or proteins), to electromagnetic media (dipoles in continuous media, orientation of domains), to interaction of living organisms \cite{Whitesides2002}. \\

{\bf Summary of the paper.} This paper provides a new geometric framework for continuum evolutionary models of particle systems whose interactions  depend on orientation. Interestingly enough, this framework is completely general and can be applied to the evolution of any geometric quantity. The problems which involve both density and orientation are perhaps the most intricate, and for pedagogical  reasons we shall treat them towards the end of the paper, after we have first illustrated the general framework by treating the familiar cases of scalars and 3-forms (particle densities). \\

After this preparation establishing the pattern for the results in simpler cases, we shall derive the covariant evolution equation (\ref{so3eq}) for quantities which depend on both particle densities and orientations. 
In this framework, one may use the same principles in deriving evolution equations for any geometric quantity. Of course, the forms of the various evolution equations will strongly depend on what geometric quantity is being considered. \\

Once the equations of motions for the geometric quantities are derived, we shall concentrate on the establishment of singular analytical solutions of these equations.  Earlier work for the case of density evolution  \cite{HoPu2005,HoPu2006}  demonstrated  that when such solutions exist, they play a dominant role in the dynamics. 
Thus, we seek additional examples of the formation of
coherent, propagating singularities in solutions of evolutionary PDEs.
These singularities arise as a result of the competition between
nonlinear and nonlocal processes in various familiar vector spaces. We
follow the same strategy as \cite{HoPu2005,HoPu2006} in taking
variations of a free energy that are induced by infinitesimal actions
of diffeos on these vector spaces. To close the equations, we also
introduce a geometric analog of Darcy's Law, based on the dual
representation of these actions. This approach yields several other
nonlinear nonlocal PDEs whose solutions form coherent propagating
singular structures in finite time.  The present approach should be useful whenever geometry figures in the formation and evolution of moving singularities. However, the development of this approach is only beginning and the full extent of its applications remains to be proven. Besides the applications mentioned earlier, see for example, the spontaneous formation of singularities (\emph{self-focusing})  in the spin wave turbulence \cite{ZaLvSt1974}. 
\rem{
{, ZaLvFa1992, Lv1994, EzRa1990} and perhaps also in game dynamics \cite{Sm1982, Ha1996, OeRi2001, EsSa2003, HoSi2003} might be of future interest for applications of this model, because of their distinct geometric properties. 
}
\\

{\bf Plan.} The plan of the paper is the following.  Section 2 reviews earlier work on the gradient flow evolution for density and formulates the geometric order parameter (GOP) equation. This formulation summons the \emph{diamond} operator from differential geometry, whose properties are discussed in Section 3.   Section 4 introduces a necessary condition for the existence of singular solutions of GOP equations in the general case. Applications require explicit calculation of the diamond operator for each geometric quantity, which is accomplished in Section 5. Section 6 derives explicit expressions for GOP equations of motion and Section 7 gives examples of singular solutions. Finally, Section 8 is devoted to the equations of motion for orientation densities whose Lie-algebra-valued singular solutions are called \emph{gyrons}. We also compare the results of our theory to recent experiments on self-organization of oriented particles. 

\section{Problem statement} 
\rem{
Suppose we dealing with particles with rotational degrees of freedom, which for example may be
suspended  in viscous fluid. This is the case with \dots  Similar models can also be used for
self-organization of biological organisms which take into account that organisms (e.g. flies
or roaches) possess a tendency to respond to a preferred direction. \\

Suppose that the macroscopic state of a system of many particles at time $t$ and position
$\bx$ is defined by a set of order parameters $\{\kappa\}=\kappa^{(p)}(\bx,t)$,
$p=1,2,3,\dots, N$, which take values in their corresponding vector spaces $V^{(p)}$,
$p=1,2,3,\dots, N$. For example, $\kappa^{(p)}$ may denote a density, or concentration, or it may
denote a vector field, tensor field, or differential form of any type. Each vector space has its
dual space, defined in terms of a pairing 
\begin{eqnarray*}
\langle \cdot\,,\, \cdot \rangle^{(p)}:\, V^{(p)}\times
{V^*}^{(p)}\mapsto\mathbb{R}
\quad\hbox{for}\quad
p=1,2,3,\dots, N,
\end{eqnarray*}
with no sum on the superscript $(p)$. In addition, we assume that the physical situation dictates a
free energy expressible in terms of the order parameters as $E[\{\kappa\}]$, where square
brackets $[\,\cdot\,]$ denotes dependence on the set of order parameters $\{\kappa\}$ which may be
spatially nonlocal, that is, may depend on their average values. \\
}

We consider continuum evolution of the macroscopic state of a
system of many particles at time $t$ and position $\bx$ that is defined
by an order parameter $\kappa(\bx,t)$, which take values in a vector
space $V$. The vector space has a dual space $V^*$, defined in terms of
the $L^2$ pairing 
\begin{eqnarray*}
\langle\boldsymbol\cdot\,,\,\boldsymbol\cdot\rangle:\, V\times{V^*}\mapsto\mathbb{R}
\,.
\end{eqnarray*}
For example, scalar functions are dual to densities, one-forms are
dual to two-forms and vector fields are dual to one-form densities. \\

In addition, we assume that the physical situation dictates a free
energy, which is a functional of the order parameter expressed as
$E[\kappa]$, where square brackets $[\,\cdot\,]$ denote dependence
which may be spatially nonlocal. That is, $E[\kappa]$ may also be a
functional of $\kappa$; for example, it could depend on the spatially averaged or filtered value of $\kappa$ defined later. Hence, the
variation of total integrated energy is given by the pairing,
\begin{eqnarray*}
\delta E[\kappa]
=
\bigg\langle 
\delta\kappa 
\,,\, 
\frac{\delta E}{\delta \kappa}
\bigg\rangle
=
\int 
\delta\kappa 
\cdot
\frac{\delta E}{\delta \kappa}
\,,
\end{eqnarray*}
where dot $(\,\cdot\,)$ denotes the appropriate pairing of vector and covector indices of $\kappa$ to
produce a density, an $n-$form (denoted as $\Lambda^n$) which then may be integrated to yield a
real number.  Of course, $n$ is the dimension of the space. In this paper, we shall concentrate on the cases $n=1,2,3$. In this setting, we seek evolution equations for the order parameter
$\kappa(\bx,t)$ that
\\(1) respect its vector space property $\kappa\in V$;
\\(2) reduce to gradient flows when $\kappa$ is a density; and 
\\(3) possess solutions that may aggregate the order parameter $\kappa$ into ``clumpons'' (quenched
states) which propagate and interact as singular weak solutions. 
\\

Thus we seek evolution equations whose solutions describe a
geometric order parameter supported on embedded subspaces of the
ambient space. For example, these solutions may be spatially
distributed on curves in 2D, or on surfaces in 3D. In fact, we seek
evolution equations for which these embedded singular solutions     are {\it attractors} which emerge even from arbitrary smooth
initial conditions, as in the case of the emergent singular densities
for the nonlocal gradient flows studied in \cite{HoPu2005,HoPu2006}. Our
ultimate goal is to find classes of equations that are relevant for the
evolution of a macroscopic order parameter that may be of potential use in the design of directed self-assembly processes in nanoscience.\\

Previous work focused on emergent singularities in an order parameter
density \cite{HoPu2005,HoPu2006}. These singularities correspond to the formation of dense clumps of maximum possible density in
 self-assembly of nano-particles. 
  Two different cases were distinguished in the model. The first case arises when the mobility of the particles $\mu$ always remains finite, in which case the evolution forces the particles to collapse into a set of $\delta$-functions. These $\delta$-functions can be understood as a large-scale view of individual clumps. The second case arises upon modeling the clumping process in greater detail, by limiting the mobility so that it will eventually decrease to zero at some maximal value of \emph{averaged} density. This leads to formation of patches of constant density. (See also \cite{Velazquez2002} for the case of density-dependent mobility.)  In the next section, we shall cast the results of previous work into the geometric framework taken here in the derivation of the equations. 

\subsection{An example: the HP equation for the gradient flow of a density}$\quad$

Holm and Putkaradze \cite{HoPu2005,HoPu2006} derived the HP gradient flow, whose singular densities (the clumpons) emerge in finite time for any smooth initial conditions possessing a maximum density. The HP equation is the gradient flow of a density $\rho\in\Lambda^n$ given by 
\begin{eqnarray}
\label{HP-eqn}
\frac{\partial \rho}{\partial t} 
=
-\,{\rm div}\, 
\bigg(
\rho \mu[\rho]\,\nabla\, \frac{\delta E}{\delta \rho}
\bigg)
\,.
\end{eqnarray}
Equation (\ref{HP-eqn}) has a clear physical meaning as follows. 
A potential at a given point $\delta E/\delta \rho$ gives rise to the force 
$ \nabla (\delta E/\delta \rho)$. The particles move with covariant velocity $\eta$ proportional to the force applied to them. The mobility $\mu$ gives the coefficient of proportionality. Thus,  the instantaneous contravariant velocity at a given point is 
\begin{eqnarray}
\eta=\left( \mu \nabla \delta E/\delta \rho\right)^\sharp. 
\label{eta-vel}
\end{eqnarray}
The familiar musical operations sharp $(\sharp)$ and flat
$(\flat)$ raise and lower vector indices, respectively, thereby mapping a covector into a contravector and vice versa, as needed for the operations of divergence, Lie derivative, etc. to make proper mathematical sense. In particular, the contravariant velocity $\mathbf{u}$ advects the particle density according to 
\begin{eqnarray}
\label{HP-geom-eqn0}
\frac{\partial \rho}{\partial t} 
=
-\,\pounds_{(\mu {\nabla}\frac{\delta E}{\delta \rho})^\sharp}
\rho
\,,
\end{eqnarray}
which provides the geometric meaning of the HP evolution equation (\ref{HP-eqn}). \\

In the traditional framework, each extensive quantity is assumed to be
conserved, so its density per unit volume $\rho$ satisfies the
continuity equation 
\begin{eqnarray}
\label{cont-eqn}
\frac{\partial \rho}{\partial t} 
=
-\,{\rm div}\, \mathbf{J}
\,,
\end{eqnarray}
where $\mathbf{J}$ is the flux density of the conserved quantity.
This flux density is assumed to be
proportional to the density of the extensive quantity. It is also
taken to be proportional to the {\it gradient} of the
thermodynamic conjugate variable $\delta E/\delta\rho$ in the First Law
for specific energy $E$, modified to be contravariant so that its
divergence may be properly defined. Thus the
name, ``gradient flow.'' Finally, a phenomenological
characterization of the freedom of motion of the particles is
introduced, called their ``mobility'' $\mu[\rho]$, which transforms as
$\rho$.  Thus, one finds the {\it Darcy Law} for flow through a porous
medium, for example, as, 
\begin{eqnarray}
\label{DarcyLaw}
\mathbf{J} 
=
\rho\,\Big(\mu\, \nabla\, \frac{\delta E}{\delta \rho}\Big)^\sharp
\,.
\end{eqnarray}
By this thermodynamic reasoning one recovers the HP equation
(\ref{HP-eqn}), from the continuity equation (\ref{cont-eqn}) and
the definition of mobility in Darcy's Law (\ref{DarcyLaw}). As a
consequence, the energy $E[\rho]$ evolves according to
\begin{eqnarray}
\label{EnergyLaw}
\frac{d E}{dt}
&=&
\bigg\langle  
\,\Big(\mu\, \nabla\, \frac{\delta E}{\delta \rho}\Big)^\sharp
,\, 
\Big(\rho \,\nabla\, \frac{\delta E}{\delta \rho}\Big)
\bigg\rangle
\nonumber \\&=&
\int
\rho\,\mu[\rho]
\Big|\nabla\, \frac{\delta E}{\delta \rho}\Big|^2
\,d\,^nx
=
\int
\frac{1}{\rho\mu[\rho]}
\big|\mathbf{J}\big|^2
\,d\,^nx
\,.
\end{eqnarray}
Note that for attracting particles, $E<0$, so the 
absolute value of energy decays in time $d |E|/dt<0$ when $\rho\,\mu[\rho]>0$. \\

Holm and Putkaradze \cite{HoPu2005,HoPu2006} analyzed the solutions of the gradient flow 
(\ref{HP-eqn}) when the mobility density $\mu[\rhobar]$ depended on the spatially averaged
density $\rhobar=H*\rho$, defined as the convolution of density $\rho$ with the kernel $H$. The
latter was chosen to be the Green's function for the Helmholtz operator. The energy density was
chosen in \cite{HoPu2005,HoPu2006} so that $\delta E/\delta \rho=G*\rho$, where $G$ was a Helmholtz
kernel whose scale length is larger than that for $H$. Thus, $\delta E/\delta \rho$ depended on the average density at the lengthscale of $G$, rather than the pointwise density.\\

With these choices, the solution
$\rho$ of the HP gradient flow, paired with a smooth scalar  test function $\phi$  satisfies 
\begin{eqnarray*}
\Big\langle \frac{\partial \rho}{\partial t}\,,\, \phi \Big\rangle
&=&
\Big\langle 
-\,{\rm div}\, 
\bigg(\rho
\Big(\mu[\rhobar] \,\nabla\, \frac{\delta E}{\delta \rho}\Big)^\sharp
\bigg)
\,,\, 
\phi
\Big\rangle
\\&=&
\Big\langle 
\rho\Big( \mu[\rhobar]\,\nabla\, \frac{\delta E}{\delta \rho}\Big)^\sharp
 \,,\,  \nabla\phi
\Big\rangle
\\&=&
\Big\langle 
\rho
\,,\, 
\Big( \mu[\rhobar]\,\nabla\, \frac{\delta E}{\delta \rho}\Big)^\sharp\cdot\nabla\phi
\Big\rangle
\\&=&
\Big\langle 
\rho
\,,\, 
\pounds_{( \mu[\rhobar]\,\nabla\, \frac{\delta E}{\delta \rho})^\sharp}\,\phi
\Big\rangle
\,.
\end{eqnarray*}
Under the the $L^2$ pairing 
$\langle\boldsymbol\cdot\,,\,\boldsymbol\cdot\rangle$, the third line matches with the derivative of a delta function; so the HP gradient flow admits singular solutions. 

\subsection{Singular solutions for the HP gradient flow of a
density}$\quad$\smallskip

Provided $\mu[\rhobar]$ and $\delta E/\delta \rho$ are sufficiently
smooth, the HP equation (\ref{HP-eqn}) admits solutions $\rho$ that are
sums of delta functions supported on $N$  manifolds of dimension $K_a$,
$a=1,2,3,\dots,N$, with coordinates $s$ embedded in $\mathbb{R}^3$.
Namely,
\begin{eqnarray}
\label{rho-weakN}
\rho(\bx,t)
&=&
\sum_a
\int _{s}
p_a(t,s)\delta\big(\bx - \bq_a(t,{s})\big)\,ds
\,,
\\
\rhobar(\bx,t)
&=&H*\rho
=
\int H(\bx,\bx')\rho(\bx')\,d\,^3x'
\nonumber\\
&=&
\sum_a
\int _{s}
p_a(t,s)H\big(\bx\,,\,\bq_a(t,{s})\big)\,ds
\,.
\label{rhobar-weakN}
\end{eqnarray}
Integrating this solution ansatz for density $\rho(\bx,t)$ against a smooth test function $\phi$
yields the dynamics of the parameters $p_a(t,s)$ (which are also densities) and the
positions $\bq_a(t,s)\in\mathbb{R}^3$, as follows. After integrating by parts, one finds
(suppressing obvious subscripts on $s_a$ and $s_b$)
\Large
\begin{eqnarray*}\label{weak-soln-eqns0}
&&\hspace{-5mm}\Big\langle\phi
\,,\,
\rho_t + \nabla\cdot\Big(\rho\,  \mu(\rhobar) \nabla (G*\rho)\Big)
\Big\rangle
=
\int\phi(\bx)
\sum_{a=1}^N \int _{s_a}
\dot{p}_a(t,s)\,\delta\big(\bx-\bq_a(t,s)\big)\,d\,^3x
\\&&\hspace{-5mm}+
\int\nabla\phi(\bx)\cdot
\sum_{a=1}^N\int_{s_a}
p_a\Big(\mathbf{\dot{q}}_a 
-
 \sum_{b=1}^N \int _{s_b}
p_b(t,s)\mu\big(\rhobar \big)
\nabla G\big(\bx\,,\,\bq_b(t,{s})\big)
\Big)\delta\big(\bx-\bq_a(t,s)\big)\,d\,^3x
\end{eqnarray*}
\LARGE
Thus, substituting the singular solution ansatz (\ref{rho-weakN}) and pairing with a smooth test function $\phi$ results in an expression in which $\phi$ and its gradient $\nabla\phi$ appear linearly. \\
 
Matching coefficients of $\phi$ and $\nabla\phi$ then yields a closed set of equations for the parameters  $p_a(t,s)$ and $\bq_a(t,s),$ $a=1,2,\dots,N,$ of the solution ansatz (\ref{rho-weakN}), as
\begin{eqnarray}
\dot{p}_a(t,s) &=& 0
\,,
\label{weak-soln-peqn}\\
\mathbf{\dot{q}}_a(t,s)  &=& 
\mu \big(\rhobar\big)
 \frac{\partial}{\partial \bq_a}
\sum_{b=1}^N \int _{s_b} p_b(s_b) G\big(\bq_a(t,{s})\,,\,\bq_b(t,{s}_b)\big)\,d{s}_b
\,.
\label{weak-soln-qeqn}
\end{eqnarray}
Here the average density $\rhobar$ is introduced in equation (\ref{rhobar-weakN}). Thus, the density weights $p_a(t,s)=p_a(s)$ are preserved, and the positions $\bq_a(t,{s})$ in
(\ref{rho-weakN}) follow the characteristics of the velocity $\eta=\mu(\rhobar)\nabla(G*\rho)$ along
the Lagrangian trajectories given by $\bx=\bq_a(t,s)$. \\

{\bf Remarks.}
\begin{itemize}
\item
The result (\ref{weak-soln-peqn}-\ref{weak-soln-qeqn}) holds in any
number of dimensions. For example, it admits singular solutions of
co-dimension one, supported on moving points in 1D, along moving curves
in 2D and on moving surfaces in 3D. 

\item
Holm and Putkaradze \cite{HoPu2005,HoPu2006} proved that the singular solutions
(\ref{rho-weakN}) emerge spontaneously from smooth initial conditions and demonstrated their
emergence in numerical simulations of equation (\ref{HP-eqn}) in one spatial dimension.  In these
1D simulations, the mass of each individual solution remained constant, as required by equation
(\ref{weak-soln-peqn}). Moreover, when two singular solutions collided, they were found to {\em add} their weights
$p_1$ and $p_2$,  thereby ``clumping'' together. Eventually, all the singular solutions
concentrated into a single ``clumpon,'' whose weight (mass) equaled the total weight of the initial
condition.

\item
The dynamics (\ref{weak-soln-peqn},\ref{weak-soln-qeqn}) for the singular solution
(\ref{rho-weakN}) of the HP equation (\ref{HP-eqn}) is a bit degenerate, because the weights
$p_a(t,s)=p_a(s)$ each turn out to be preserved. This occurs because substituting the singular
solution into the HP equation for density produces only one term proportional to the test
function, $\phi$, which in turn yields trivial dynamics of the weights $p_a$, $a=1,2,\dots,N$. 
The other terms are proportional to $\nabla\phi$ and determine the dynamics of $\bq_a(t,s)$.

\item
The general situation for the variational evolution of an arbitrary order parameter $\kappa\in V$
(not just a density $\rho\in\Lambda^n$) might be expected to possess multiple terms proportional
to both $\phi$ and $\nabla\phi$, and thereby produce nontrivial dynamics for both $p_a(t,s)\in V$
and $\bq_a(t,s)\in\mathbb{R}^3$. Note, however, that the singular solution (\ref{rho-weakN}) of HP
only existed, because when paired with the test function $\phi$ the substitution of
(\ref{rho-weakN}) into (\ref{HP-eqn}) produced no higher derivatives than $\nabla\phi$. This was
the key condition for possessing singular solutions of HP.
\end{itemize}

\subsection{Geometric order parameter equations}

$\quad$\smallskip$\quad$

{\bfi Cautionary note about signs:} From now on, we will use the convention that the velocity of particles moving under the potential $E$
is $\mathbf{v}=-\mu \nabla \delta E/\delta \rho$. This convention  introduces a minus sign, so that the force will be defined as \emph{minus} the gradient of the potential. Hence, the formulas for Darcy velocity in the previous section each acquire a ``$-$" sign. 
The equations for continuous evolution in \cite{HoPu2005,HoPu2006} may be obtained by changing the sign of energy $E \rightarrow -E$ in the final formulas. 
\\[2mm]

{\bfi Geometry of order parameters.} Order parameters for continua need not be densities. Instead, they may be amplitudes, or phases, or direction fields, or any other type of geometrical object,  such as scalars, vectors, tensors, differential one-forms, two-forms, etc. Such geometrical objects are characterized by how they transform under smooth invertible maps with smooth inverses (diffeomorphisms, called diffeos for short). The corresponding infinitesimal transformations of geometrical objects are defined as their Lie derivatives with respect to smooth vector fields. \\

\rem{
In seeking phenomenological evolution equations for geometric order parameters, we shall begin by focusing on their {\em reversible deterministic} aspects. The irreversible or stochastic aspects of these evolution laws such as diffusion will be added elsewhere. Thus, to begin, geometrical quantities should evolve by Lie derivative. 
}

To create evolution equations for these quantities, we mimic the pattern of the gradient-flow equation (\ref{HP-eqn}). 
Two features of this equation guide its generalization: \\

\rem{ 
First, the Lie derivative acts on an object that transforms the same way as the order parameter does. In the gradient-flow equation for the density (\ref{HP-eqn}), the divergence is recognized as the Lie derivative of the mobility $\mu[\rhobar]$, which also transforms as a density. \\
} 

First, we seek an evolution equation for a geometric quantity $\kappa$ based on the following conservation law in weak form,
\begin{eqnarray}
\label{HP-geom-eqn}
\Big\langle \frac{\partial \kappa}{\partial t}\,,\, \phi \Big\rangle
=
\Big\langle
-\,\pounds_{u[\kappa]}
\kappa
\,,\, \phi \Big\rangle
\,,
\end{eqnarray}
In other words, the geometric quantity $\kappa$ is Lie-dragged with velocity $u[\kappa]$,  a vector field whose dependence on $\kappa$ must be defined as a suitable generalization of Darcy's velocity. 

Second, the vector field $u$ which Lie-drags that object must be bilinear in the mobility 
$\mu$  and the variation of energy $\delta E/\delta \kappa$. In the gradient-flow equation (\ref{HP-eqn}), the Darcy vector field is $-(\mu[\kappa] \,\nabla\, \frac{\delta E}{\delta \rho})^\sharp$. We seek the generalization of this vector field for an arbitrary geometrical order parameter.\\

The previous derivation of the HP equation  \cite{HoPu2005,HoPu2006} must be
modified to admit an arbitrary geometric object $\kappa\in V$ in any
representation vector space $V$ of the diffeomorphisms, rather than
specializing to densities. 
 Physically, this corresponds to identifying the order parameter by
how it transforms under the diffeos and specifying its flux vector
accordingly. This approach will allow us to address the formation of
singularities in order parameters that are not densities.
In particular, we compute the geometric variational flow for an order
parameter $\kappa\in V$ taking values in any vector space $V$, by
following steps similar to those followed in \cite{HoPu2005,HoPu2006} for deriving HP. Namely,
\begin{eqnarray}
\Big\langle \frac{\partial \kappa}{\partial t} \,,\, {\phi} \Big\rangle
&=&
\Big\langle 
\delta\kappa 
\,,\, 
\frac{\delta E}{\delta \kappa}
\Big\rangle
\nonumber \\&=&
\Big\langle 
\frac{\delta E}{\delta \kappa}
\,,\, 
-\,\pounds_{u({\phi})}\,\mu[\kappa]
\Big\rangle
\nonumber \\&=&-\,
\Big\langle  
\mu \,\diamond\, \frac{\delta E}{\delta \kappa}
\,,\,
u({\phi}) 
\Big\rangle
\nonumber \\&=&
\Big\langle  
\Big(\mu \,\diamond\, \frac{\delta E}{\delta \kappa}\Big)
,\, 
({\phi}\,\diamond\,\kappa\,)^\sharp
\Big\rangle
\label{key-formula} \\&=&
\Big\langle  
({\phi}\,\diamond\,\kappa\,)
,\, 
\Big(\mu \,\diamond\, \frac{\delta E}{\delta \kappa}\Big)^\sharp
\Big\rangle
\nonumber \\
\Big\langle \frac{\partial \kappa}{\partial t} \,,\, {\phi} \Big\rangle
&=&
\Big\langle  -\,
\pounds_{(\mu\, {\diamond}\frac{\delta E}{\delta \kappa})^\sharp}
\kappa,\,{\phi}
\Big\rangle
\label{GOP-eqn-deriv}
\end{eqnarray}
The first step invokes symmetry of the $L^2$ pairing and takes
variations of the order parameter $\kappa\in V$ by using the Lie derivative with respect to the vector field $u({\phi})$. The second step introduces the {\bfi diamond operation}
$\diamond$. The diamond operation is defined in terms of the Lie derivative $\pounds_u$ action of a vector field $u$ acting on variables $a\in V$ and $b\in V^*$ that are dual under the $L^2$ pairing by
\begin{eqnarray}
\langle \pounds_u a \,,\, b \rangle
=
\langle u \,,\, a \diamond b \rangle
=
-\,\langle  a \,,\, \pounds_u b \rangle
=
-\,\langle u \,,\, b \diamond a \rangle
\label{diamond-def}
\end{eqnarray}
As before, the sharp $(\sharp)$ operation raises vector indices in the key third step (\ref{key-formula}), which defines the velocity vector field 
$u({\phi})=-\,({\phi}\,\diamond\, \kappa\,)^\sharp$ in terms of
$\kappa\in V$, $\phi \in V^*$ and the diamond operation.
This is the generalization of vector field $\eta$ in equation (\ref{eta-vel}) from a density to an arbitrary vector quantity,
\begin{eqnarray}\label{Darcy-kappa}
{u}({\phi})=-\,({\phi}\,\diamond\,\kappa\,)^\sharp
\,.
\end{eqnarray} 
The last step used by  (\ref{GOP-eqn-deriv}) is the definition of the diamond operation to express the Lie derivative of the mobility
$\pounds_u \kappa$   with respect to the
vector field 
\begin{equation} 
u=(\mu \, {\diamond}\,\frac{\delta E}{\delta \kappa})^\sharp. \label{Darcy-vel-general} 
\end{equation} 
From the physical point of view, the vector field velocity defined by  (\ref{Darcy-vel-general})  provides the natural generalization we seek of  the Darcy velocity $\mu \nabla \delta E/\delta \rho$. 
Thus we obtain the following {\bfi geometric order parameter (GOP)} equation,
\begin{eqnarray}
\label{GOP-eqn}
\frac{\partial \kappa}{\partial t}
=
-\,\pounds_{(\mu \, {\diamond}\frac{\delta E}{\delta \kappa})^\sharp}
\kappa
\,.
\end{eqnarray}
When the order parameter is a density
$\kappa=\rho\in\Lambda^n$, then diamond specializes to gradient, the Lie
derivative becomes a divergence and one recovers the HP equation of
\cite{HoPu2005,HoPu2006}.  Thus, the GOP equation (\ref{GOP-eqn})
generalizes the concept of gradient flow of a density to ``diamond
flow'' of any geometric quantity. The corresponding energy equation
follows from (\ref{GOP-eqn}) as
\begin{eqnarray}
\frac{d E}{dt}
=
\Big\langle \frac{\partial \kappa}{\partial t} \,,\, 
\frac{\delta E}{\delta \kappa} \Big\rangle
&=&
\left\langle  -\,
\pounds_{(\mu\, {\diamond}\frac{\delta E}{\delta \kappa})^\sharp}
\kappa,\,\frac{\delta E}{\delta \kappa}
\right\rangle
\nonumber \\
&=&
-\,
\left\langle  
\Big(\mu \,\diamond\, \frac{\delta E}{\delta \kappa}\Big)
,\, 
\Big(\kappa \,\diamond\,\frac{\delta E}{\delta \kappa}\,\Big)^\sharp
\right\rangle
\label{GOP-erg}
\end{eqnarray}
In what follows, we shall consider
several other forms of this equation for geometric order parameters in
various vector spaces.  The explicit forms of these GOP equations
and their corresponding energetics need to be calculated using the
definition of diamond and its properties.


\section{Properties of the diamond operation}

The {\bfi diamond operation} $\diamond$ is defined in (\ref{diamond-def}) for Lie
derivative $\pounds_\eta$ acting on dual variables $a\in V$ and $b\in V^*$ by
\begin{equation}
\langle b \diamond a \,,\, \eta \rangle
\equiv
-\,
\langle b \,,\, \pounds_\eta\, a \rangle
=:
-\,
\langle b \,,\, a\, \eta \rangle
\,,
\end{equation}
where Lie derivative with respect to right action of the diffeomorphisms on elements of $V$ is also denoted by concatenation on the right.  
The diamond operator takes two dual quantities $a$ and $b$ and produces a quantity  dual to a vector field, \emph{i.e.}, a $a\diamond b$ is a one-form density. For example, as we compute in more details below, if $a$ is a scalar and $b=B \mbox{d}^n \mathbf{x}$ is an $n$ form (where $n$ is the dimension of the space), then 
\[ 
a  \diamond  b =   \left( B \nabla a\right)  \cdot \rm{d} \mathbf{x} \otimes \rm{d}^n \mathbf{x}
\] 
The $\diamond$ operation is also known as the ``dual representation'' of this right action of the Lie algebra of vector fields  on the representation space $V$ \cite{HoMaRa1998}.  \\

When paired with a vector field $\eta$, the diamond operation has the following three useful properties:
\begin{enumerate}
\item
It is antisymmetric
\[
\langle \, b \diamond a
+
a \diamond b\,,\, \eta \rangle
=
0
\,.
\]
\item
It satisfies the product rule for Lie derivative
\[
\langle \, \pounds_\xi \,(b \diamond a)\,,\, \eta \, \rangle
=
\langle\, (\pounds_\xi\,b) \diamond a
+
b \diamond (\pounds_\xi\,a)\,,\, \eta \, \rangle.
\]
\item
It is antisymmetric under integration by parts
\[
\langle \,
db \diamond a
+
b \diamond da\,,\, \eta 
\, \rangle
=
0
\,.
\]
\end{enumerate}
\bigskip

First, the $\diamond$ operation is {\bfi antisymmetric},
\begin{equation}\label{diamond-skew}
\langle b \diamond a \,,\, \eta \rangle
=
-\,
\langle a \diamond b\,,\, \eta \rangle
\,,
\end{equation}
as obtained by using,
\begin{equation}\label{antisym-rel}
\langle b \,,\, \pounds_\eta\, a \rangle
+
\langle \pounds_\eta\, b \,,\, a \rangle
=
0
\,,\quad\hbox{or,}\quad
\langle b \,,\, a\,\eta \rangle
+
\langle b\,\eta \,,\, a \rangle
=
0
\,,
\end{equation}
and the symmetry of the pairing
$\langle\boldsymbol\cdot\,,\,\boldsymbol\cdot\rangle$. \\

{\bf Example.} For example, if $a=\mathbf{A}\cdot\nabla$ and 
$\eta=\boldsymbol{\eta}\cdot\nabla$ are vector fields and 
$b=\mathbf{B}\cdot d\mathbf{x}\otimes d^nx$
is a one-form density in Euclidean coordinate notation, 
we find for the $L^2$ pairing in one dimension
\[\langle\,a\,,\,b\,\rangle = \int a\contract b 
= \int (\mathbf{A} \cdot \mathbf{B}) d^nx\]
In this case, integrating the product rule identity for contraction ($\contract$)
\[
\pounds_\eta (a\contract b)
=
(\pounds_\eta a)\contract b + a \contract \pounds_\eta b 
\]
yields, for the density (3-form) $\pounds_\eta (a\contract b)$ with homogeneous boundary
conditions, 
\begin{eqnarray*}
\int \pounds_\eta (a\contract b)
&=&
\int {\rm div}\,\big((\mathbf{A} \cdot \mathbf{B})\,\boldsymbol{\eta}\big)\,d\,^3x = 0
\\&=&
\int (\pounds_\eta a)\contract b + a \contract \pounds_\eta b 
\\&=&
\langle \pounds_\eta  a \,,\, b \rangle
+
\langle a \,,\, \pounds_\eta b \rangle
\\&=&
\langle b  \,,\, \pounds_\eta  a\rangle
+
\langle a \,,\, \pounds_\eta b \rangle
=
-\,\langle b \diamond a \,,\, \eta \rangle
-\,
\langle a \diamond b\,,\, \eta \rangle
\end{eqnarray*}
As expected, (\ref{diamond-skew}) follows for this case, re-affirming that the diamond operation is
skew-symmetric.  
In other notation, the intermediate equation may be rewritten as
\begin{eqnarray*}
0
=
\langle \pounds_\eta  a \,,\, b \rangle
+
\langle a \,,\, \pounds_\eta b \rangle
&=&
\langle \,{\rm ad}_\eta a \,,\, b \rangle
+
\langle a \,,\, {\rm ad}^*_\eta b \rangle
\,,
\end{eqnarray*}
where $\pounds_\eta b =  {\rm ad}^*_\eta b $ for a one-form density $b$.
In vector notation, 
\[
\pounds_\eta  a
=
{\rm ad}_\eta a 
=
[\eta \,,\, a]
=
 \boldsymbol{\eta}\cdot\nabla\mathbf{A}
-\, \mathbf{A}\cdot\nabla\boldsymbol{\eta}
\]
and the intermediate equation is verified by integration by parts.

Second, the $\diamond$ operation satisfies the \emph{ product rule under the Lie derivative},
\begin{equation}
\langle \pounds_\xi\,(b \diamond a)\,,\, \eta \rangle
=
\langle (\pounds_\xi\,b) \diamond a\,,\, \eta \rangle
+
\langle b \diamond (\pounds_\xi\,a)\,,\, \eta \rangle.
\end{equation}
This property may be verified directly, as
\begin{align}
\langle \pounds_\xi\, b \diamond a\,,\, \eta \rangle
&+
\langle b \diamond \pounds_\xi\, a\,,\, \eta \rangle
=
\langle b\,\xi\,\eta \,,\, a \rangle
-
\langle b\,\eta\,\xi \,,\, a \rangle
\nonumber\\
&=
\langle a \,,\, b\,({\rm ad}_\xi\,\eta) \rangle
=
-\, \langle a \diamond b\,,\,({\rm ad}_\xi\,\eta) \rangle
\nonumber\\
&=
-\, \langle {\rm ad}^*_\xi (a \diamond b)\,,\,\eta \rangle
=
-\, \langle \pounds_\xi (a \diamond b)\,,\,\eta \rangle
=
   \langle \pounds_\xi (b \diamond a)\,,\,\eta \rangle
\,,
\end{align}
upon using $\langle b\,\xi \,,\, a\,\eta \rangle
+
\langle b\,\xi\eta \,,\, a \rangle = 0$, implied by (\ref{antisym-rel}),
in the first step. \\

Finally,  the $\diamond$ operation is {\bfi antisymmetric under integration by parts},
\begin{equation}
\langle \, b \diamond da \,,\, \eta \rangle
=
-\,\langle db \diamond a\,,\, \eta \rangle
\,,
\end{equation}
as obtained from commutation of the two types of derivative 
$(da)\eta = d(a\eta)$ and integration by parts,
\begin{equation}
\langle \,b \,,\, d(a\eta)\, \rangle
+
\langle \,db \,,\, a\eta\, \rangle
=
0
\,.
\end{equation}
These three properties of $\diamond$ are useful in computing the
explicit forms of the various geometric gradient flows for order
parameters (\ref{GOP-eqn}). Of course, when the order parameter is a
density undergoing a gradient flow, then one recovers HP from
(\ref{GOP-eqn}).

\section{Dissipative bracket and its properties}

The formula for energy decay in (\ref{GOP-erg}) suggests the following bracket notation for the time derivative of a functional $F[\kappa]$,
\begin{eqnarray}
\frac{d F[\kappa]}{dt}
=
\Big\langle \frac{\partial \kappa}{\partial t} \,,\, 
\frac{\delta F}{\delta \kappa} \Big\rangle
&=&
\left\langle  -\,
\pounds_{(\mu \, {\diamond}\frac{\delta E}{\delta \kappa})^\sharp}
\kappa \,,\,
\frac{\delta F}{\delta \kappa}
\right\rangle
\nonumber \\
&=&
-\,
\left\langle  
\Big(\mu \,\diamond\, \frac{\delta E}{\delta \kappa}\Big)
,\, 
\Big(\kappa\,\diamond\,\frac{\delta F}{\delta \kappa}\,\Big)^\sharp
\right\rangle
\nonumber \\
&=:&
\{\!\{\,E\,,\,F\,\}\!\} [\kappa]
\label{bracketdef}
\end{eqnarray}
Thus, the GOP equation (\ref{GOP-eqn}) may be written in bracket form as
\begin{eqnarray}
\label{GOP-eqn-brkt}
\frac{\partial \kappa}{\partial t}
=
-\,\pounds_{(\mu\, {\diamond}\frac{\delta E}{\delta \kappa})^\sharp}
\kappa
=
\{\!\{\,E\,,\,\kappa\,\}\!\}
\,.
\end{eqnarray}

The properties of the GOP brackets $ \{\!\{\,E\,,\,F\,\}\!\}$ defined in equation (\ref{bracketdef}) are determined by the diamond operation and the choice of the mobility
 $\mu[\kappa]$. For example, the GOP bracket in equation (\ref{bracketdef}) is  neither symmetric nor antisymmetric, for a general choice of mobility $\mu[\kappa]$. 
However, any physical choice of mobility should produce strict dissipation of energy, \emph{i.e.}
 \begin{equation}
  \{\!\{\,E\,,\,E\,\}\!\} \leq 0 
  \label{Energy}
  \end{equation} 
  One particular choice for the mobility satisfying this energy dissipation condition  is $\mu[\kappa]=\kappa M[\kappa]$, where $M[\kappa] \geq 0$ is a non-negative scalar functional of $\kappa$. For this choice, one finds 
  \[ 
    \{\!\{\,E\,,\,E\,\}\!\} =-M[\kappa] \bigg\langle 
 \kappa\diamond \frac{\delta E}{\delta \kappa}
   \,,\,
 \Big(\kappa \diamond \frac{\delta E}{\delta \kappa} \Big)^\sharp
 \bigg\rangle
\leq 0
 \]

\subsection{Properties of the GOP Bracket: Leibnitz identity} 

\begin{proposition}[Leibnitz property]
The GOP bracket (\ref{bracketdef}) satisfies the Leibnitz property \cite{OrPla2003}. That is, it satisfies 
\[
\{ \{ \,EF\,,\,G\,\} \}[\kappa] =F \{ \{ \,E\,,\,G\,\} \}[\kappa] +E\{ \{ \,F\,,\,G\,\} \}[\kappa] 
\] 
for any functionals $E,F$ and $G$ of $\kappa$. 
\end{proposition}
\begin{proof}
For arbitrary scalar functionals  $E$ and $F$ of $\kappa$ and any smooth vector field $\eta$, the Leibnitz property for the functional derivative and for the Lie derivative  together imply
\begin{eqnarray*} 
\bigg\langle  
\mu \diamond 
\left( 
\frac{\delta (EF )}{\delta \kappa} 
\right) \, , \,  
\eta
\bigg\rangle 
&=&
\bigg\langle  
\mu \diamond 
\left( 
E \frac{\delta F}{\delta \kappa}
+
F \frac{\delta E}{\delta \kappa} 
\right) 
\, , \,  
\eta
\bigg\rangle 
\\ 
\\
&=&
\bigg\langle  \mu \, , \,  
-\pounds_\eta 
\left( 
E \frac{\delta F}{\delta \kappa}
+
F \frac{\delta E}{\delta \kappa} 
\right)
\bigg\rangle  
\\
\\
&=&
E \bigg\langle  \mu \, , \,  
-\pounds_\eta 
\frac{\delta F}{\delta \kappa}
\bigg\rangle  
+
F
\bigg\langle  \mu \, , \,   
-\pounds_\eta 
\frac{\delta E}{\delta \kappa} 
\bigg\rangle  
\\
&=&
E\bigg\langle  
\mu \diamond 
\frac{\delta F}{\delta \kappa}
\, , \,  
\eta
\bigg\rangle
+
F\bigg\langle  
\mu \diamond 
\frac{\delta E}{\delta \kappa}
\, , \,  
\eta
\bigg\rangle
\end{eqnarray*}
Choosing $\eta= \Big(\kappa\diamond \frac{\delta G}{\delta \kappa} \Big)^\sharp$ then proves the proposition that the bracket (\ref{bracketdef}) is Leibnitz. 

\end{proof}

\subsection{ Connection to Riemannian geometry} 
Following \cite{Otto2001}, we use the GOP bracket to introduce a metric tensor on the manifold connecting a ``vector'' $\partial_t \kappa$ and ``co-vector'' $\delta E/\delta \kappa$. 
That is, we express the evolution equation (\ref{GOP-eqn}) in the weak form  as 
\begin{equation} 
\bigg \langle 
\frac{\partial \kappa}{\partial t} \, , \, \psi
\bigg \rangle
=g_\kappa \left(  
\frac{\delta E}{\delta \kappa} \, , \, \psi 
\right) 
\label{Otto-def} 
\end{equation} 
for an arbitrary element $\psi$ of the space dual to the $\kappa$ space, and where 
$g_\kappa(\cdot \, , \, \cdot)$ is a symmetric positive definite function -- metric tensor -- defined on vectors from the dual space. 
(Here, we choose a $+$ sign in front of metric tensor to be consistent with the choice of the sign for energy functional throughout this paper). 

Let us first notice that 
for any choice of mobility producing a symmetric bracket (in particular, $\mu[\kappa]=\kappa M[\kappa]$) 
 \[
 \{\!\{\,E\,,\,F\,\}\!\} = \{\!\{\,F\,,\,E\,\}\!\}
 \,,
 \]
 we may regard that symmetric bracket as defining an inner product between the functional derivatives,
 \begin{equation} 
 \{\!\{\,E\,,\,F\,\}\!\} =: g_\kappa \Big(
  \frac{\delta E}{\delta \kappa} \,,\,
\frac{\delta F}{\delta \kappa}
\Big) =
\bigg \langle 
\mu \diamond \frac{\delta E}{\delta \kappa} 
\, , \, 
\left( \kappa \diamond \frac{\delta F}{\delta \kappa} \right)^\sharp 
\bigg \rangle 
 \label{gtensor} 
 \end{equation} 
 Alternatively, (\ref{gtensor}) can be understood as a symmetric positive definite function of two 
 elements of dual space $\phi,\psi$: 
 \begin{equation} 
 g_\kappa \Big(
\phi \,,\,
\psi
\Big)
= 
\bigg \langle 
\mu \diamond \phi 
\, , \, 
\left( \kappa \diamond \psi \right)^\sharp 
\bigg \rangle . 
 \label{gtensor2} 
 \end{equation} 
 Notice that $g(\phi,\phi) \geq 0$ by (\ref{Energy}). 
 
 \begin{proposition}[Metric property of the GOP equation]
For the choice of metric tensor (\ref{gtensor},\ref{gtensor2}), the GOP equation (\ref{GOP-eqn}) may be expressed as the metric relation (\ref{Otto-def}). 
 \end{proposition}
\begin{proof}
 For an arbitrary  element $\psi$ in the dual space,
\begin{eqnarray}
\Big\langle \frac{\partial \kappa}{\partial t} \,,\, {\psi} \Big\rangle
&=&
\Big\langle 
- \pounds_{(\mu\, {\diamond}\frac{\delta E}{\delta \kappa})^\sharp}
\kappa
\, ,\,
{\psi}
\Big\rangle
\nonumber \\
&=&
\Big\langle 
\kappa \diamond \psi  
\, , \, 
\left( \mu \diamond \frac{\delta E}{\delta \kappa} \right)^\sharp 
\Big\rangle
\nonumber \\
&=&
\Big\langle 
\mu \diamond \frac{\delta E}{\delta \kappa} 
\, , \, 
\left( \kappa \diamond \psi  \right)^\sharp 
\Big\rangle =
 g_\kappa \left( \frac{\delta E}{\delta \kappa} \, , \, \psi \right) 
\label{Ottoderiv} 
\end{eqnarray}
\end{proof}
This approach harnesses the powerful machinery of Riemannian geometry to the mathematical analysis of the GOP equation (\ref{GOP-eqn}). This opens a wealth of possibilities, but it also limits the analysis to mobilities $\mu$ for which the GOP bracket (\ref{bracketdef}) is symmetric and positive definite, as in the modeling choice  $\mu[\kappa]=\kappa F[\kappa]$.
Nevertheless, its strength seems to outweigh its limitations so we plan to apply this approach for modeling physical processes using GOP equations in our future work. 

\section{Existence of singular solutions for the GOP equation (\ref{GOP-eqn})} 
Let's choose free energy and mobility $E[\kappa],\,\mu[\kappa]\in H^2$  and $\kappa\in H^{-1}$
(which includes delta functions). For example, one may choose $\delta E/\delta \kappa = G*\kappa$
and mobility tensor $\mu[\kappabar]$ with $\kappabar = H*\kappa$ with Helmholtz kernels $G$ and $H$
with two different length scales, as done for the HP equation  (\ref{HP-eqn}) from (\ref{HP-eqn})
in \cite{HoPu2005,HoPu2006}.  \\

The geometric order parameter equation (\ref{GOP-eqn}) is then,
\begin{eqnarray*}
\frac{\partial \kappa}{\partial t}
=
-\,\pounds_{(\mu[\kappabar]\, {\diamond}\frac{\delta E}{\delta \kappa})^\sharp}
\kappa
\equiv
-\,\pounds_{u[\kappa]} \kappa
\,,\quad\hbox{with}\quad
u[\kappa]
\equiv
\Big(\mu[\kappabar]\, {\diamond}\frac{\delta E}{\delta \kappa}\Big)^\sharp
\,.
\end{eqnarray*}

When paired with a smooth test function $\phi$ dual to $\kappa$ in $L^2$, the solution $\kappa$
satisfies, cf. equations (\ref{key-formula}, \ref{GOP-eqn-deriv})
\begin{eqnarray}
\Big\langle \phi\,,\,\frac{\partial \kappa}{\partial t} \Big\rangle
&=&
\Big\langle \phi\,,\, 
-\,\pounds_{u[\kappa]} \kappa
\Big\rangle
\nonumber \\
&=&
\Big\langle \phi\diamond \kappa\,,\, u[\kappa]\Big\rangle
\nonumber \\
&=&
\Big\langle  \kappa  \,,\,\pounds_{u[\kappa]} \phi \Big\rangle
\nonumber \\
\label{key-formula-again} 
\Big\langle \phi\,,\,\frac{\partial \kappa}{\partial t} \Big\rangle
&=&
\bigg\langle  \kappa\, 
  \,,\, 
-\,\pounds_{\big(\mu \diamond \frac{\delta E}{\delta \kappa} \big)^\sharp} 
\phi \bigg\rangle
\label{kappa-pairing}
\end{eqnarray}
One may now substitute the singular solution Ansatz for GOP 
\begin{eqnarray}
\label{kappa-weakN}
\kappa(\bx,t)
&=&
\sum_a
\int _{s}
p_a(t,s)\delta\big(\bx - \bq_a(t,{s})\big)\,ds
\,,
\end{eqnarray}
with $\kappa,\,p_a\in V$  into the left hand side
of (\ref{kappa-pairing}). This substitution will produce only terms proportional to $\phi$ and $\nabla\phi$, after the appropriate integrations by parts. Fortunately,  this precisely matches the corresponding terms on the right side for any $\kappa$, since the Lie derivative for arbitrary tensor quantity $\phi$ only contains terms proportional to $\phi$ and its gradient.

\section{Two familiar examples} 

We write the GOP equation (\ref{GOP-eqn}) explicitly in two familiar cases. In each case we express the GOP equation of motion (\ref{GOP-eqn}) in terms of the Leibnitz bracket (\ref{bracketdef}) as 
\begin{eqnarray}
\frac{d F}{d t}
&=&
\{\!\{\,E\,,\,F\,\}\!\} [\kappa]
\label{Lbracket-eqn}
\end{eqnarray}
for any smooth functional $F[\kappa]$.\\

In three dimensional Euclidean coordinates, the Lie derivatives of a scalar $f$ and a density $D\,d^3x$ are:
\begin{eqnarray}
-\pounds_\mathbf{v}\ f
= -\mathbf{v}\cdot\nabla\,f
\quad\hbox{and}\quad
-\pounds_\mathbf{v}\,(D\,d^3x)
= -\nabla\cdot(D\mathbf{v})\ d^3x
\,.
\label{w-eqn}
\end{eqnarray}
Having these two fomulas, we may compute from the definition of diamond (\ref{diamond-def}) that
\begin{eqnarray}
\mu[f] \diamond \frac{\delta E}{\delta f} 
=
\frac{\delta E}{\delta f}\nabla \mu
\quad\hbox{and}\quad
\mu[D] \diamond \frac{\delta E}{\delta D} 
&=&
-\,\mu[D] \nabla \frac{\delta E}{\delta D}
\end{eqnarray}
\subsection{Functions (bottom forms)}
For functions $f$, the relation (\ref{key-formula-again}) with diamond becomes,
\begin{eqnarray}
\bigg\langle  
\mu[f] \diamond \frac{\delta E}{\delta f}
  \,,\,
\big(\phi\diamond f \big)^\sharp 
\bigg\rangle
&=&-\,
\bigg\langle
\Big(\frac{\delta E}{\delta f}\nabla \mu[f]\Big)
\,,\,  (\phi\nabla f)^\sharp
\bigg\rangle
\label{f-eqn}
\end{eqnarray}
Consequently, the {\bfi Leibnitz bracket} (\ref{bracketdef}) for functionals of a scalar function $f$ is 
\[ 
\{ \{ \,E\,,\,F\,\} \}[f] 
= 
-
 \int
 \frac{\delta E}{\delta f}
 \frac{\delta F}{\delta f}  
  \nabla f  \cdot \nabla \mu[f] 
 \mbox{d}^3 \mathbf{x}
\] 
As expected, this Leibnitz bracket is not symmetric for an arbitrary choice of mobility $\mu[f]$. However, it produces proper energy dissipation $\{\!\{\,E\,,\,E\,\}\!\} \leq 0$ provided  $\nabla f  \cdot \nabla \mu[f] \ge0$.

\subsection{Densities (top forms)}
For densities, the diamond pairing is
\begin{eqnarray}
\bigg\langle  
\mu \diamond \frac{\delta E}{\delta D}
  \,,\,
\big(D \diamond \phi\big)^\sharp 
\bigg\rangle
&=&
\bigg\langle
\mu \nabla \frac{\delta E}{\delta D}
\cdot 
(D\, \nabla \phi)^\sharp
\bigg\rangle
\end{eqnarray}
and the Leibnitz bracket (\ref{bracketdef}) is expressed as
\begin{equation} 
\{ \{ \,E\,,\,F\,\} \}[D] = 
 -\int 
 D \mu[D] 
 \nabla  \frac{\delta E}{\delta D} 
 \cdot 
\nabla  \frac{\delta F}{\delta D}  
  \mbox{d}^3 \mathbf{x}
  \label{3formdissip0} 
\end{equation}
In this case
\begin{eqnarray}
\frac{dE}{dt}
&=&
\{ \{ \,E\,,\,E\,\} \}[D] 
\nonumber\\
&=&
- \, 
 \int 
 D \mu[D] \,
 \Big|\nabla  \frac{\delta E}{\delta D} \Big|^2
   \mbox{d}^3 \mathbf{x}
  \nonumber\\
&\le&
 0 
 \quad\hbox{provided}\quad
 D \mu[D] \ge 0
  \label{3formdissip} 
\end{eqnarray}

\paragraph{\bfi Remark.}
These relations determine the dependence of the GOP energetics
(\ref{GOP-erg}) on the choice of mobility in other geometric cases case, simply by replacing $\phi$ in fomula (\ref{gtensor} ) above by the appropriate variational derivative of energy. 

\section{GOP equations of motion} 

The diamond relations above lead to the following equations of motion.\\

\subsection{GOP equation for a scalar} 
First, for a scalar $f\in\Lambda^0$
\begin{eqnarray}
\label{scalareq} 
\frac{\partial f}{\partial t} 
&=& -\,\pounds_{(\mu[f] \diamond\frac{\delta E}{\delta f})^\sharp} f 
\\
&=& -\,\pounds_{(\frac{\delta E}{\delta f}\nabla  \mu[f]  )^\sharp} f 
\\
&=& -\,\Big(\frac{\delta E}{\delta f}\nabla \mu[f] \Big)^\sharp\cdot\nabla f
\end{eqnarray}

Equation (\ref{scalareq}) can be rewritten in {\bfi characteristic form} as $f(x,t)=$ const along 
\[ 
\frac{dx}{dt}=\frac{\delta E}{\delta f} \nabla \mu[f] 
\] 
This is an unusual characteristic equation, since the speed of characteristics depends 
on the nonlocal quantities $\delta E/\delta f$ and mobility $\mu$.  
\subsection{GOP equation for a density} 
 
For the density $D\,d\,^3x\in\Lambda^3$, we have
\begin{eqnarray}
\frac{\partial D }{\partial t}
&=& -\,\pounds_{(\mu[D]\diamond\frac{\delta E}{\delta D})^\sharp} D
\nonumber\\
&=& {\rm div}\,\bigg(D \Big( \mu[D]\nabla\frac{\delta E}{\delta D}\Big)^\sharp\bigg)
\,.
\label{HP-sign}
\end{eqnarray}
This recovers the HP equation (\ref{HP-eqn}). 
  (apart from the minus sign 
in front due to the revised definition of velocity, as noted in the beginning of section 2.3). 
\section{Examples of singular solutions for the GOP equation (\ref{GOP-eqn})} 

Substitution of the singular solution for GOP 
\begin{eqnarray}
\label{kappa-weakN-again}
\kappa(\bx,t)
&=&
\sum_a
\int _{s}
p_a(t,s)\delta\big(\bx - \bq_a(t,{s})\big)\,ds
\,,
\end{eqnarray}
into both sides of (\ref{key-formula-again}) will produce only terms proportional to $\phi$ and $\nabla\phi$, after
the appropriate integrations by parts. Matching corresponding terms yields the dynamics of
$p_a(t,s)$ and $\bq_a(t,{s})$. \\

For example, by GOP  formulas (\ref{GOP-eqn},\ref{kappa-pairing},\ref{f-eqn}), the scalar $f$ satisfies,
\begin{eqnarray}
\nonumber
\Big\langle \phi\,,\,\frac{\partial f}{\partial t} \Big\rangle
&=&
\bigg\langle  f\, {\diamond}\frac{\delta E}{\delta f}
  \,,\,\big(\phi\diamond \mu[f]\big)^\sharp \bigg\rangle
\\
\nonumber
&=&-\,
\bigg\langle
\Big(\frac{\delta E}{\delta f}\nabla f\Big)
\,,\,  (\phi\nabla \mu[f])^\sharp
\bigg\rangle
\\&=&
\nonumber
\bigg\langle
f
\,,\, 
{\rm div}\,\phi\,
\Big(\frac{\delta E}{\delta f}
\nabla \mu[f]\Big)^\sharp
\bigg\rangle
\\&=&
\bigg\langle
f
\,,\, 
\nabla\phi\,\cdot\,
\Big(\frac{\delta E}{\delta f}
\nabla \mu[f]\Big)^\sharp
+\,
\phi\
{\rm div}\,
\Big(\frac{\delta E}{\delta f}
\nabla \mu[f]\Big)^\sharp
\bigg\rangle
\label{f-eqn1}
\end{eqnarray}
One inserts
\begin{eqnarray}
\label{f-weakN}
f(\bx,t)
&=&
\sum_a
\int _{s}
p_a(t,s)\delta\big(\bx - \bq_a(t,{s})\big)\,ds
\,,
\end{eqnarray}
with scalars $\,p_a$ and $a=1,2,\dots,N$ 
into the left hand side of this equation and matches terms to find the equations for 
the parameters $\,p_a$ and $\bq_a$ of the singular scalar equation. These take the forms,
\begin{eqnarray}
\dot{p}_a(t,s) 
&=& 
{p}_a(t,s) 
\,{\rm div}\,
\Big(\frac{\delta E}{\delta f}
\nabla \mu[f]\Big)^\sharp
\bigg|_{\bx = \bq_a(t,{s})}
\label{weak-fsoln-peqn}\\
{p}_a(t,s) \,\mathbf{\dot{q}}_a(t,s)  
&=&
{p}_a(t,s)\, 
\Big(\frac{\delta E}{\delta f}
\nabla \mu[f]\Big)^\sharp
\bigg|_{\bx = \bq_a(t,{s})}
\label{weak-fsoln-qeqn}
\end{eqnarray}
One then chooses $E[f]$ and $\mu[f]$ so that the right hand sides make
sense as functional relations. \\

In Figure~\ref{fig:scalarevolution}, we demonstrate the spatio-temporal evolution of a scalar given by (\ref{scalareq}). We have taken $\delta E/\delta f=H*f$ where $H$ is the inverse Helmholtz operator $H(x)=e^{-|x|/\alpha}$ with length-scale, or filter width $\alpha=1$. The initial conditions consists of equally spaced $\delta$-peaks in $f$ with random strengths $p_a$:  $f(x,0)=\sum_a p_a(0) \delta(x-q_a(0))$, $-1/2<p_a(0)<1/2$. The  evolution of the positions and amplitudes of the peaks is highly complex and shows strong sensitivity to the initial conditions. Thus, although the exact solution to the nonlocal PDE follows from the system of ordinary differential equations (\ref{weak-fsoln-peqn},\ref{weak-fsoln-qeqn}), the solution of that system may still be highly complex and sensitive to the initial conditions of the $p$'s and $q$'s. Fortunately, an exact analytical solution for this system of ODEs is available in the important case of evolution of a pair of $\delta$-peaks. This solution imparts general understanding of the long-term behavior of  (\ref{weak-fsoln-peqn},\ref{weak-fsoln-qeqn}) so we shall describe it in detail. 
\remfigure{ 
\begin{figure} [h]

\centering 

\includegraphics[width=16cm]{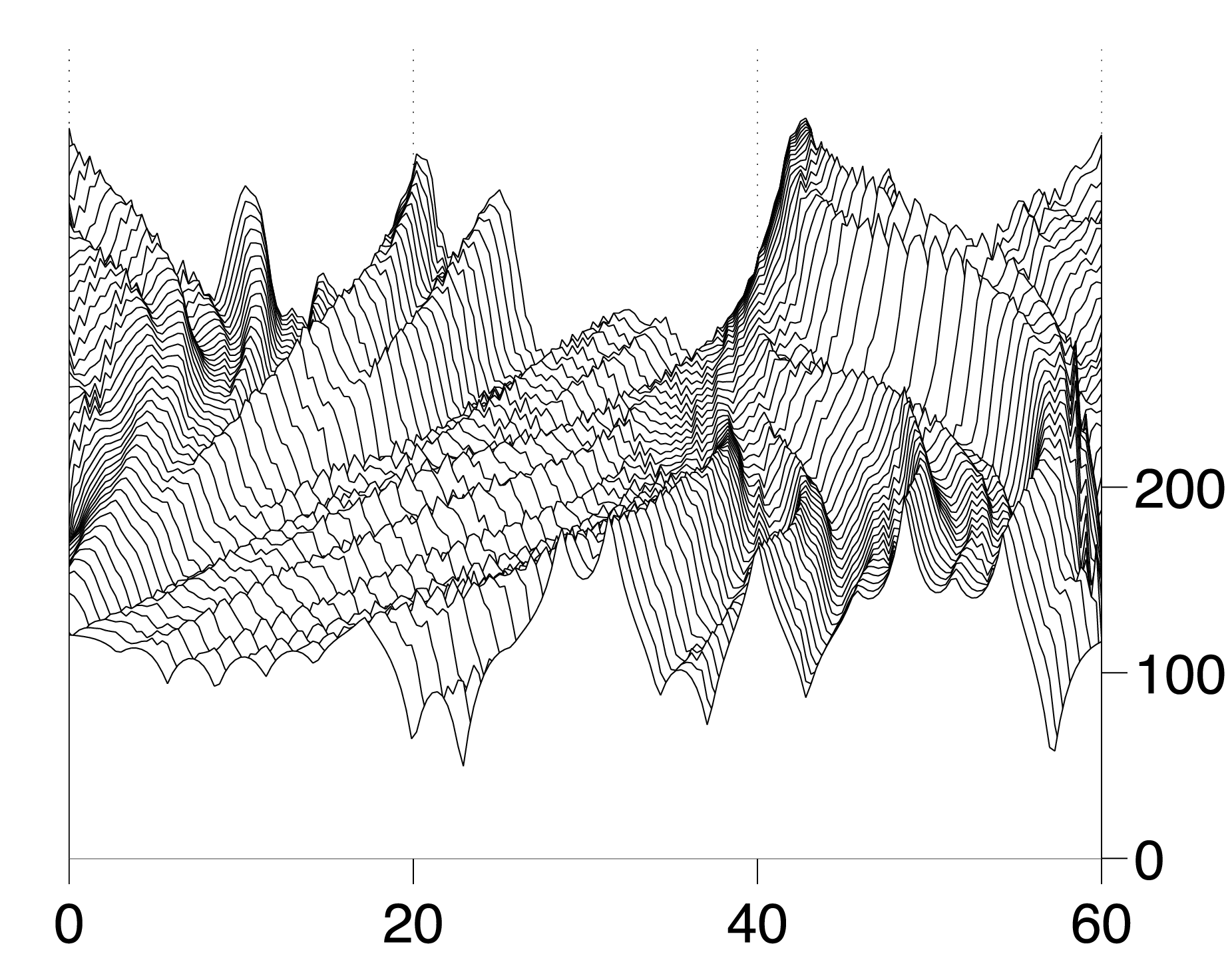} 
 \caption{  Numerical simulation demonstrates the 
evolution of an initial set of $\delta$-functions in $f$. The vertical coordinate represents  $\overline{f}=H*f$, which remains finite even when $f$ forms $\delta$-functions for our choice  $H(x)=e^{-|x|}$. This simulation of the initial value problem
for equation (\ref{scalareq}) uses generalized initial conditions 
$f(x,0)=\sum_a p_a(0) \delta(x-q_a(0))$ with random initial amplitudes $-1/2<p_a(0)<1/2$. 
 } 
 \label{fig:scalarevolution} 
 \end{figure} 
 } 

\bigskip

We assume initial conditions for the scalar $f$ as
\[
f(x,0)=p_0 \delta(x-q_0)+s\, p_0 \delta(x-q_0)
\,.
\]
 Here, the sign $s=+1$ specifies symmetric initial conditions, whereas $s=-1$ enforces antisymmetric initial conditions.  
We shall choose the sign $s$ so the evolution due to (\ref{weak-fsoln-peqn},\ref{weak-fsoln-qeqn}) preserves the symmetric position of the $\delta$-function at $x= \pm q(t)$. 
If this choice is possible, then a solution of $p(t),q(t)$ in terms of quadratures can be found. Not all energies $E[f]$ and mobilities $\mu[f]$ allow symmetry preservation; however, many physically relevant cases do have this property. By reflection symmetry of the PDE, the even and odd parts of the solution are separately invariant. Consequently, we may assume a solution \emph{for all} $t$ in the form, 
\begin{equation} 
\label{symsoln}
f(x,t)=p(t) \Big( \delta \left(x-q(t) \right)+s\, \delta \left(x+q(t) \right) \Big)
\,.
\end{equation}
Hence, the \emph{four} equations (\ref{weak-fsoln-peqn},\ref{weak-fsoln-qeqn}) reduce by reflection symmetry to only \emph{two} equations for $p(t),q(t)$: 
\begin{eqnarray}
\dot{p}(t) 
&=& 
p(t) 
\,{\rm div}\,
\Big(\frac{\delta E}{\delta f}
\nabla \mu[f]\Big)^\sharp
\bigg|_{x = \pm q(t)}
\label{peqn2}\\
\dot{q}(t)  
&=&
\Big(\frac{\delta E}{\delta f}
\nabla \mu[f]\Big)^\sharp
\bigg|_{x = q_a(t)}
\label{qeqn2}
\end{eqnarray}
Let us first notice that the derivative is taken at  $x=\pm q$, then 
\[ 
\,{\rm div}\,
\Big(\frac{\delta E}{\delta f}
\nabla \mu[f]\Big)^\sharp
\bigg|_{x = \pm q(t)}=2 \frac{\partial}{\partial q} \left( \Big(\frac{\delta E}{\delta f}
\nabla \mu[f]\Big)^\sharp 
\bigg|_{x = \pm q(t)} \right) 
\] 
If we now denote 
\begin{equation} 
\label{Psipqdef} 
\Psi(p,q)=\left( \Big(\frac{\delta E}{\delta f}
\nabla \mu[f]\Big)^\sharp 
\bigg|_{x = \pm q(t)} \right), 
\end{equation} 
then equations (\ref{peqn2},\ref{qeqn2}) can be written as 
\begin{equation} 
\label{pqeqn2} 
\dot{p}=2 p \frac{\partial}{\partial q} \Psi(p,q)  
\,,\hskip1cm 
\dot{q}=\Psi(p,q) 
\,.
\end{equation}
We are now ready to prove the following. \\

{\bfi Theorem 2} \\ 
{\em  Suppose $\Psi(p,q)$ defined by (\ref{Psipqdef}) is homogeneous with $\Psi(p,q)=p^\gamma \psi(q)$, and  $s$ (symmetry parameter) is chosen so the solution retains its symmetry under the evolution. Then, a set of initial conditions $f=p_0 \left(\delta(x-q_0)+s\delta(x+q_0) \right)$ exists, such that $q(t)$ in the solution (\ref{symsoln}) collapses $q\to0$ in finite time, whenever $\psi(q)$ is a continuous function bounded away from zero with $|\psi(q)| \geq A>0$ in some neighborhood of $q=0$. 
}
 \\[4mm] 
\emph{Note.} Many physical choices of energy and mobility admit the required homogeneity of $\Psi(p,q)$. For example \cite{HoPu2005,HoPu2006} selected $\delta E/\delta f=G*f$, $\mu=1-H*f$, where $G$ and $H$ are given functions. 
This choice of energy and mobility implies $\Psi(p,q)=p^2 \psi(q)$, so $\gamma=2$. 
Another example we shall employ here is $G(x)=e^{-x^2/\alpha^2}$ with $\alpha$ being a fixed parameter, and $\nabla \mu=1$, since the explicit formulas are particularly simple. This choice yields $\gamma=1$.  \\ [4mm] 
\emph{Proof of Theorem 2} \\ 
First, let us notice that equations (\ref{pqeqn2}) integrate exactly (in terms of quadratures) if we assume that $\Psi(p,q)=p^\gamma \psi(q)$. Indeed, (\ref{pqeqn2}) 
are equivalent to 
\begin{equation} 
\label{pqeqn3} 
\frac{dp}{dq}=2 p\frac{1}{\psi(q)}  \frac{d}{d q} \psi(q) 
\end{equation}
which integrates exactly in terms of initial conditions $p(0)=p_0$, $q(0)=q_0$ as 
$$
p=p_0 \left( \frac{\psi(q)}{\psi(q_0)}  \right)^2.  
$$
Notice that by the assumption $|\psi(q)| \geq A>0$ in some neighborhood of $q=0$ we can choose $q_0$ sufficiently close to $0$ so $\psi(q_0) \neq 0$. Then, the $q$-equation of (\ref{pqeqn2}) gives 
\begin{equation} 
\label{qeqn4} 
\dot{q}=p_0 \psi(q_0)^{-2 \gamma} \psi(q)^{2 \gamma+1}  
\end{equation} 
If, by assumption,  $\psi(q)<-A<0$ if $|q|<\delta$ for some $A>0,\delta>0$, let us choose $p_0>0$ and $0<q_0<\delta$.  Then 
$$
\dot{q} \leq  -\frac{ p_0 }{ \psi(q_0)^{-2 \gamma}} A^{2 \gamma+1}  
$$ 
so $q(t)$ goes to zero in time $t_*$ not exceeding 
\begin{equation} 
\label{tstar} 
t_* \leq \frac{q_0  \psi(q_0)^{2 \gamma}}{p_0 A^{2 \gamma+1} } 
\end{equation} 
The case $\psi(q) \geq A>0$ is treated analogously by choosing $p_0<0$. 
\\[4mm] 
 We performed a simulation of the evolution of a scalar  with initial conditions of the type (\ref{kappa-weakN-again}). More precisely, we chose $\delta E/ \delta f=G*f$ with $G(x)=e^{-x^2\alpha^2}$, and $\nabla \mu=1$. This gave values $\gamma=1$, $\psi(q)=e^{-4 q^2/\alpha^2}$, so $\psi$ satisfies the conditions of the theorem. We chose initial conditions $f=p_0 (\delta(x-q_0)-\delta(x+q_0))$, since in this case, the antisymmetric solutions preserve symmetry under evolution, with the choice $p_0=-1/8$, $q_0=5/3$. \\

The evolution of positions for $\delta$-functions for this simulation is shown  in Figure~\ref{fig:characteristics}. The exact solution is shown with a solid line. The dashed line illustrates the position according to the numerics. The exact solution collapses in finite time whereas the numerical solution shows exponentially slow approach of the $\delta$-peaks. We attribute the apparent discrepancy for large time to numerical dissipation. Indeed, introducing a term $-\epsilon p$ mimicking numerical dissipation in the right-hand side of the $p$ equation 
in (\ref{pqeqn2}) prevents solution collapse, for any $\epsilon>0$, however small $\epsilon$ may be. Since every numerical scheme must necessarily involve some numerical dissipation or distortion, we believe it may be difficult to construct a numerical scheme that shows exact collapse of the solutions. This numerical question will be addressed elsewhere in future studies. \\
\remfigure{ 
\begin{figure} [h]

\centering 
\includegraphics[width=16cm]{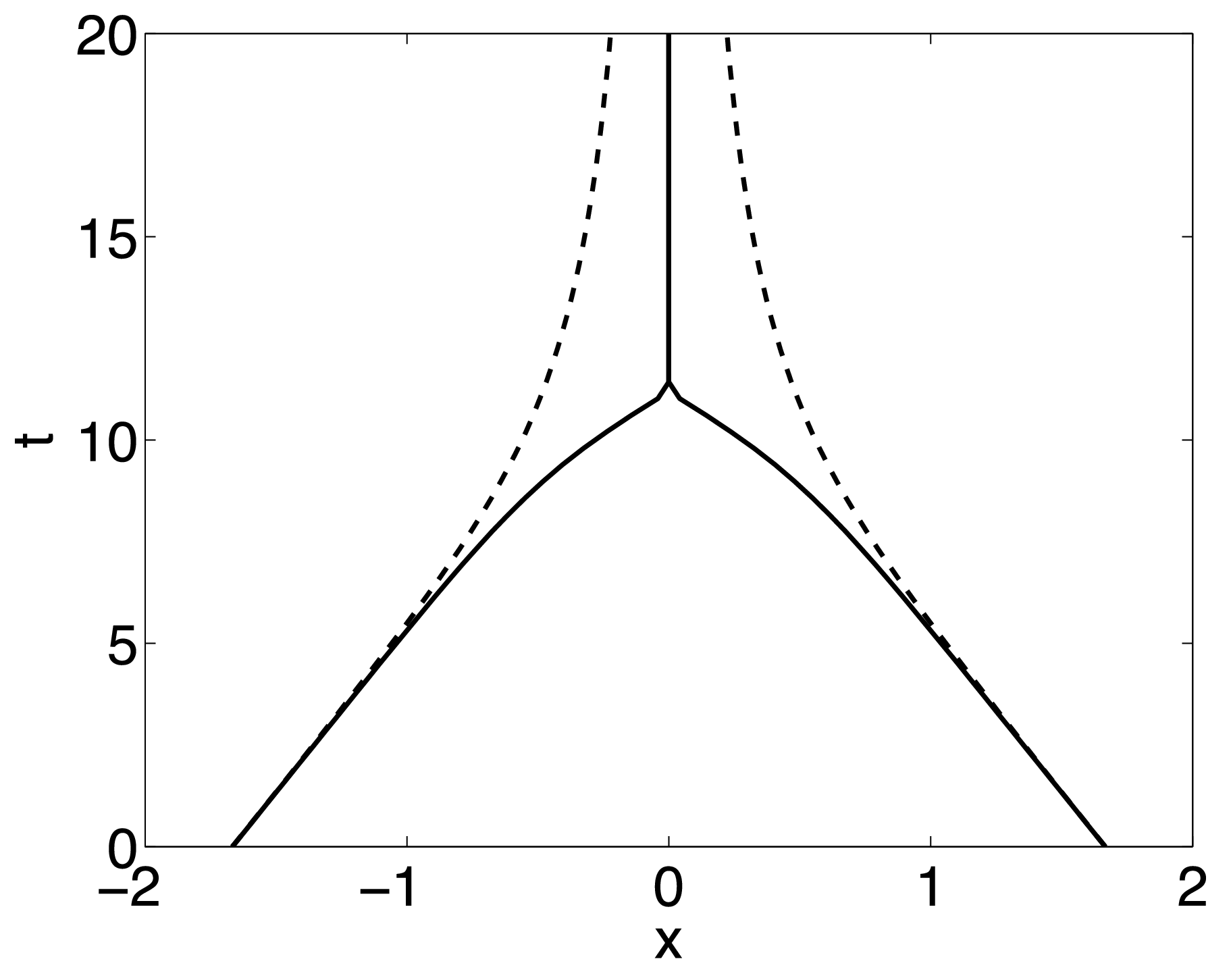}
 \caption{ 
 Collapse of a solution with initial condition having two $\delta$-functions with equal strength, opposite in sign $f(x,0)=p_0 \left(  \delta(x-q_0) -\delta(x+q_0) \right)$ with $p_0=1/8$, $q_0=2$. Exact solution is shown with a  solid line, results of numerics is given with a dashed line. 
 } 
 \label{fig:characteristics} 
 \end{figure} 
} 

The solution behavior may be investigated further by comparing the numerical and exact solutions for amplitude (\ref{weak-fsoln-peqn}) in the evolution of a single $\delta$-peak. For this comparison, one may start with a single delta-function $f(x,0)=p_0 \delta(x)$ and choose parameters
$\mu=1-H*f$, $\delta E/\delta f=G*f$ and $G(x)=e{-|x|}$, $H(x)=\frac{1}{2}
e^{-|x|}$. In this case, equation (\ref{weak-fsoln-peqn}) reduces to $\dot{p} =-2 p^3$, whose solution is 
\begin{equation} 
\label{ampsol} 
\frac{1}{p^2}=\frac{1}{p_0^2}+4 t
\end{equation} 
We illustrate the validity of this prediction in Fig.~\ref{fig:ampevolution}. 
\\[4mm]
\remfigure{ 
\begin{figure} [h]

\centering 
\includegraphics[width=6in]{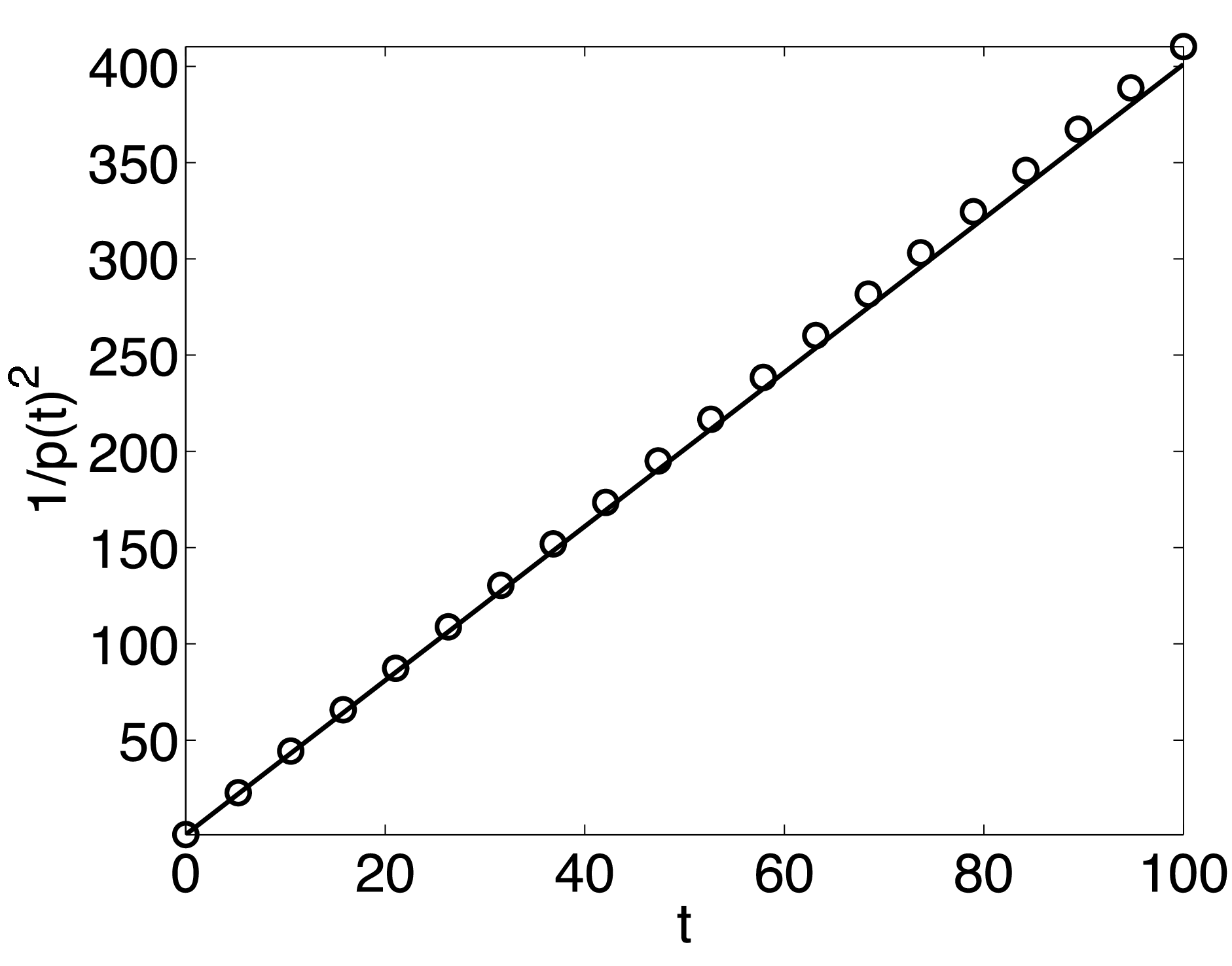} 
 \caption{ 
Evolution of the $1/p(t)^2$ versus time (circles).
The theoretical prediction (\protect{\ref{ampsol}}) is shown as a solid line. 
Note that  (\protect{\ref{ampsol}}) contains no fitting parameters. 
 } 
 \label{fig:ampevolution} 
 \end{figure} 
}

The equations for the singular-solution parameters $\,p_a$ and
$\bq_a$ for the other quantities in equation (\ref{f-weakN}) may be
found the same way, by substituting the solution  ansatz  (\ref{f-weakN}) above into the formulas (\ref{GOP-eqn},\ref{kappa-pairing},\ref{f-eqn1}) and  matching terms.  \\

Not unexpectedly, the density case recovers previous results for the HP equation. 

\section{ $so(2)$- and  $so(3)$-valued densities and gyrons} 
\subsection{Derivation of equations}

An interesting situation for practical applications occurs in the  
evolution of particles whose mutual interaction depends on their relative orientation. One example familiar from everyday life is the attraction of floating particles of non-circular shape, such as squares or stars. In this situation, the attraction between any two particles arising through their mutual deformation of the surface depends on their relative orientation. \\

Let us consider the motion of a set of particles whose energy $E$ depends on their relative orientations. A reasonable physical assumption is that each particle carries its own measure of orientation. (This is the case with floating particles -- stars and squares, say.) However, notice that the rotation angle and density are independent -- one may find dense regions with little relative rotation and regions of small mass density but large relative rotation. This possibility leads us to study the evolution of a physical quantity $\kappa$ which carries mass density and orientation separately. 
Mathematically, this corresponds to a density (an element of $\Lambda^3$) that carries a ``charge'' taking values in the real numbers $\mathbb{R}$ for the mass and in the dual of the Lie algebra $so(n)^*$ with $n=2$ or $3$ for orientation in two or three dimensions. Such a quantity is written as $\kappa \in ( \mathbb{R} \otimes so(n)^*) \otimes\Lambda^3 $. These are (dual) Lie-algebra-valued ``charge'' densities.  Let $e_a$ with $a=1,2,\dots,n(n-1)/2$ be a set of basis vectors for the Lie algebra $so(n)$.  These basis vectors satisfy
\begin{equation} 
\label{struct-consts} 
\left[ e_b, e_c \right]=t^d_{bc} e_d  \hskip1cm (b,c,d>0)
\,,
\end{equation} 
and $t^d_{bc}$ are the structure constants of the Lie algebra $so(n)$. For the case $n=3$ one recovers the familiar result that $t^d_{bc}=\epsilon_{ab}^c$ is the completely antisymmetric tensor density.
Let us expand $\kappa$ in terms of the dual basis elements $e^a$ satisfying $\langle\,e_a\,,\,e^b\,\rangle=\delta^b_a$ in the form 
\begin{equation} 
\label{rhoso3} 
\kappa \,\mbox{d}^n x
=
 \kappa_a e^a \mbox{d}^nx 
 \,.
\end{equation} 
Here, the summation on the Lie algebra index $a$ ranges from $zero$ (for the mass density) to $dim(so(n))=n(n-1)/2$ for the orientation degrees of freedom. Thus,  $a=0$ corresponds to the real-valued mass density and $a=1,2,\dots,n(n-1)/2$ denote the basis vectors of $so(n)$.  
\\

The Lie algebra $so(3)$ has three basis vectors. In contrast, $so(2)$
has only one basis vector, since only one number -- angle of rotation --
is sufficient to describe the rotation in two dimensions. Then, for
example, in three dimensions we have $\kappa \in (\mathbb{R} \otimes so(3)^*) \otimes\Lambda^3$. Correspondingly, the energy variation is given by $\delta E / \delta \kappa \in \mathbb{R} \otimes so(3)$ and is a function that takes real values for $a=0$ and takes values in the Lie algebra for $a>0$. 
The distinction between $so(3)$ and its dual $so(3)^*$ is purely
notational for the present case, since these may both be identified with $\mathbb{R}^3$. However, we shall write the formulas below in notation that would be valid for an arbitrary Lie algebra.
\\

The evolution equation for $\kappa \in (\mathbb{R} \otimes so(3)^*) \otimes\Lambda^3$  in the space of densities taking values in the dual Lie algebra $so(3)^*$ is expressed as a pairing with  a Lie algebra valued function $\phi \in \Lambda^0 \otimes
so(3)$. The mobility 
$\mu \in (\mathbb{R} \otimes so(3)^*) \otimes\Lambda^3$ lies in the same space as 
$\kappa$. Let us introduce the real-valued velocity vector 
$\boldsymbol{\eta} = -\kappa_b \mathbf{D}^\sharp \phi^b$, where the spatial covariant derivative operator $\mathbf{D}$ is expressed in components as, 
\begin{equation}
\label{spatial-covarder}
D_j
=
\partial_j 
+
{\rm ad}^*_{A_j} 
\,,\quad\hbox{or}\quad
(D_j\phi)^d
=
(\delta^d_a\partial_j
+
 t^d_{ab}A_j^b) \phi^a
 \,.
\end{equation}
Here, one sums over Lie algebra index $b$ and the operation  $(\,\cdot\,)^\sharp$ in 
$ \boldsymbol{\eta} = -\kappa_b \mathbf{D}^\sharp \phi^b$ 
raises the spatial index. This vector field has components
\begin{equation}
\label{Darcy-vel}
\eta^j
= -\kappa_b (D^j \phi)^b
:=-\,
\kappa_b\Big( \frac{\partial }{\partial x^j} \phi^b\Big)^\sharp 
+
\kappa_a t_{cb}^a A^{jc} \phi^b
\,.
\end{equation}
Thus, $\boldsymbol{\eta} $ is a real vector field in $\mathbb{R}^n$.
This formula expresses Darcy law velocity for Lie-algebra-valued quantities by using the spatial derivative in covariant form, with Lie-algebra-valued connection $\mathbf{A}^ce_c$ whose $j-$th spatial component is ${A_j}^ce_c$, with $j=1,2,\dots,n$ and $({A_j}^ce_c)^\sharp=A^{jc}e_c$. \\

Likewise, we introduce the covariant time evolution operator as,
\begin{equation}
\label{time-covarder}
D^*_t \kappa 
=
\partial_t \kappa+ {\rm ad}^*_{A_0} \kappa
=
(\partial_t \kappa_a + t^d_{ab}A^b_0 \kappa_d)e^a
\,,
\end{equation}
where $A_0=A^a_0e_a\in so(3)$ is the temporal component of the Lie algebra valued connection. We also introduce the covariant  spatial divergence, cf. equation (\ref{spatial-covarder}), as
\begin{equation}
\label{covardiverg}
\mbox{Div}^* \kappa \boldsymbol{\eta} 
=
D_j (\kappa \eta^j)
=
\partial_j (\kappa \eta^j)
+
{\rm ad}^*_{A_j} \kappa \eta^j
\,.
\end{equation}
Finally, we are ready to compute the covariant evolution equation. Applying the previous definitions and integrating by parts yields
\begin{eqnarray*} 
\left< {\rm D}_t^*\kappa\,,\, \phi \right> 
&=&
\left< 
\delta\kappa
,\, 
\frac{\delta E}{\delta \kappa}  
\right>  
:=
\left< 
-\,
{\rm Div}^* \mu[\kappa]\boldsymbol{\eta} 
,\, 
\frac{\delta E}{\delta \kappa}  
\right>  
=
\left< 
\mu[\kappa] \boldsymbol{\eta} 
,\, 
\mathbf{D}\,\frac{\delta E}{\delta \kappa}  
\right> 
\\
&=&
\left< 
\mu[\kappa]( -\kappa_b \mathbf{D}^\sharp \phi^b)
,\, 
\mathbf{D}\,\frac{\delta E}{\delta \kappa}  
\right>  
=
\left< 
{\rm Div}^*
\left(   
\kappa \Big( \mu[\kappa]_a \mathbf{D}^\sharp \frac{\delta E}{\delta \kappa_a}
\Big)
\right) 
,\,
\phi \right> 
\,. 
\end{eqnarray*} 
Thus, we find the {\it same} equation we would have derived from
thermodynamics, now written in terms of Lie-algebraic covariant derivatives,
\begin{equation} 
\label{so3eq} 
 {\rm D}_t^*\kappa
=
{\rm Div}^*
\left(   
\kappa 
\Big(\mu[\kappa]_a  \mathbf{D}^\sharp \frac{\delta E}{\delta \kappa_a}
\Big)
\right)
\,.
\end{equation} 
As one might have expected, this is the covariant form of the HP equation (\ref{HP-sign}) for the orientation ``charge density'' $\kappa=\kappa_be^b$ taking values in the dual of the Lie algebra. This is the covariant evolution equation we sought for  interaction energies that depend on both particle densities and orientations.  \\

{\bf The covariant form of Lie-algebra-valued scalar GOP equation.}
Comparing the patterns of equations   (\ref{HP-sign}) and (\ref{so3eq})  with that for the scalar case (\ref{scalareq}), we may immediately write the corresponding dynamics for a Lie-algebra-valued scalar $f=f^ae_a$ in covariant form, as
\begin{eqnarray}
\label{Lie-scalar-eq} 
D_t f
&=&  -\,\Big(\frac{\delta E}{\delta f}\,\mathbf{D} \mu[f] \Big)^\sharp
\cdot\mathbf{D} f
\,.
\end{eqnarray}
The index notation for the spatial and temporal covariant derivatives of densities in the dual Lie algebra was explained, respectively, in equations (\ref{spatial-covarder}) and  (\ref{time-covarder}). The scalar equation (\ref{Lie-scalar-eq}) may also be expressed in index notation by raising or lowering the Lie algebra indices appropriately. This observation also sets the pattern for generalizing other GOP flows to Lie-algebra values. \\

{\bf Particular modeling choices for $SO(3)$.} The Lie-algebra-valued connections $A_0$ and $\mathbf{A}$ are not yet defined. One may identify a good candidate for $A_0$ by examining the limit when the mobility vanishes,  $\mu[\kappa] =0$. In this limit, when we choose the Lie algebra $so(3)$, and make a particular choice of $A_0$, equation (\ref{so3eq})  reduces to the famous Landau-Lifshitz equation for spin waves,
\begin{equation} 
\label{LL-eqn}
  {\rm D}_t^*\kappa
=
(\partial_t \kappa_a - \epsilon^d_{ab}A^b_0 \kappa_d)e^a
=
0
\,,
\end{equation} 
or, in obvious vector notation,
\begin{equation} 
\label{LL-eqn-vec}
\partial_t \kappa - A_0 \times  \kappa
=
0
\,.
\end{equation} 
The Landau-Lifshitz equation now emerges upon making the choice $A_0={\delta E/\delta \kappa}$, so that
\begin{equation} 
\label{LL-eqn-vec2}
\partial_t \kappa 
=
\frac{\delta E}{\delta \kappa}  \times \kappa 
=
\kappa  \times \Delta\kappa
\,,\quad\hbox{when}\quad
\frac{\delta E}{\delta \kappa}  = -\, \Delta\kappa 
\,.
\end{equation} 
It remains to choose the spatial connection $\mathbf{A}$ as our final modeling parameter. For the sake of simplicity, in the rest of the paper we shall make the choice,  
\[
\mathbf{A}\cdot\mbox{d}\mathbf{x}=d\chi\chi^{-1}\in so(3)
\]
with $\chi$ in the order parameter group $SO(3)$. This choice is a ``pure gauge'' connection whose spatial curvature is always zero. The quantity $\chi\in SO(3)$ is an {\em auxiliary variable} defined by relating its advective time derivative to the temporal part of the connection $A_0$ as
\begin{eqnarray*} 
\dot{\chi}:=
\frac{\partial \chi}{\partial t} 
+ \mathbf{u}\cdot\nabla \chi
= A_0 \chi
\quad\hbox{where}\quad
 \mathbf{u} = \Big(\kappa\diamond\frac{\delta E}{\delta\kappa}\Big)^\sharp
 \quad\hbox{and}\quad
 A_0 = \frac{\delta E}{\delta\kappa}
 \,.
\end{eqnarray*} 
From its definition, one finds an evolution equation for the spatial part of the connection form, 
\begin{eqnarray} 
(d\chi\,\chi^{-1})\dot{\,}
=
(\mathbf{A}\cdot\mbox{d}\mathbf{x})\dot{\,}
&=&
\big( \partial_t \mathbf{A} 
+ \mathbf{u}\cdot\nabla \mathbf{A} 
+ (\nabla \mathbf{u)^T} \cdot \mathbf{A} \big)
\cdot\mbox{d}\mathbf{x}
\nonumber\\
&=&
dA_0 + {\rm ad}_{A_0} (\mathbf{A}\cdot\mbox{d}\mathbf{x})
\nonumber\\
&=&
\big(\partial_j A_0^a +\epsilon^a_{bc} A_0^b{A}_j^c\big)e_a 
\mbox{d}x^j
\label{Aconnection-eqn}
\end{eqnarray} 
where $A_0=A^b_0e_b\in so(3)$ is the temporal component of the Lie algebra valued connection. This formula closes the system.  Given the  choice $A_0={\delta E/\delta \kappa}$ and the relation $A_0=\dot{\chi}\chi^{-1}$ for the auxiliary variable $\chi$, this calculation of the evolution equation for $\mathbf{A}\cdot\mbox{d}\mathbf{x}=d\chi\chi^{-1}$ extends to any order parameter group. The resulting system is 
\begin{eqnarray} 
 {\rm D}_t^*\kappa
&=&
{\rm Div}^*
\left(   
\mu[\overline{\kappa}] 
\Big(\kappa_a \mathbf{D}^\sharp \frac{\delta E}{\delta \kappa_a}
\Big)
\right)
\\
\partial_t \mathbf{A} 
&+& \mathbf{u}\cdot\nabla \mathbf{A} 
+ (\nabla \mathbf{u)^T} \cdot \mathbf{A}
=
\nabla A_0 + {\rm ad}_{A_0} \mathbf{A}
\label{kA-system-eqn}
\end{eqnarray} 
where
$ \mathbf{u} = \Big(\kappa\diamond\frac{\delta E}{\delta\kappa}\Big)^\sharp$ and $ A_0 = \frac{\delta E}{\delta\kappa}$.
In general, if the connection terms are small compared to spatial 
gradients, (\ref{kA-system-eqn}) becomes simply a nonlinear diffusion equation. Alternatively, 
if the connection terms are much larger than the spatial gradients, right-hand side of (\ref{kA-system-eqn}) becomes an integrable double-bracket equation considered in \cite{BlBrRa1992}). The left hand side of (\ref{kA-system-eqn}) drives the evolution of $\kappa$ through the presence of time derivative and Landau-Lifshitz terms. 

\subsection{Measure-valued solutions: \emph{gyrons}} 
A special case that warrants further investigation arises when the Lie algebra in question is abelian, so that the structure constants vanish. As we shall show below, this case yields singular solutions. 

Let us multiply (\ref{so3eq}) by an arbitrary function $\phi \in \mathbb{R}\otimes so(3)$. After the first integration by parts, we find 
\begin{equation} 
\label{so3phi} 
\left<  \frac{\partial \kappa_b}{\partial t}, \phi^b \right> 
= \left<    \kappa_b 
\bigg(\mu[\kappa]_a \frac{\partial}{\partial x^j} 
\frac{\delta E}{\delta \kappa_a}  
\bigg)^\sharp
,\, \frac{\partial \phi^b}{\partial x_j} 
\right> 
\,.
\end{equation} 
By proceeding analogously to the calculation of HP clumpons in previous
sections, we see that singular solutions of the form
\begin{equation} 
\label{kappaansatz}
\kappa_a = \int p_a (s,t) \delta (\mathbf{x}-\mathbf{q}(s,t)) \mbox{d}s
\end{equation} 
are admitted, in which $p_a$ takes values in the dual Lie algebra. Since the right hand side of (\ref{so3phi}) contains no
terms proportional to $\phi^a$, we find that 
$\partial_t p_a(s,t)=0$, for $a=0,1,\ldots,dim(so(n))$. 
Equations for the $j$-th component of $\mathbf{q}$ which are expressed from the equation for component $b$ are: 
\begin{equation} 
\label{qeqturn} 
p_b \frac{\partial q^j}{\partial t}=p_b 
\bigg(\mu[\kappa]_a \rm{D}_j 
\frac{\delta E}{\delta \kappa_a} \Big|_{( \mathbf{x}=\mathbf{q})}
\bigg)^\sharp. 
\end{equation} 
Assuming that $p_b \neq 0$, we get simple evolution equations for the strength $p$ and coordinates $q_j$: 
\begin{equation} 
\label{pqeqnfinal} 
 \frac{\partial p_a}{\partial t} =0 \hskip1cm 
 \frac{\partial q^j}{\partial t}= 
\bigg(\mu[\kappa]_a \rm{D}_j 
\frac{\delta E}{\delta \kappa_a} \Big|_{( \mathbf{x}=\mathbf{q})}
\bigg)^\sharp. 
\end{equation} 
As we shall see, the singular solutions (gyrons) do appear in experiments. Thus, the 
singular solutions derived here are an essential part of the dynamics for particles with orientation.


\subsection{Comparison with experiments for the case of floating five-point stars } 
illustrate the power of our theory and connect it with the original problem of self-assembly of particles with orientation,  we present a numerical simulation of the system (\ref{kA-system-eqn}) formulated for the particular case of a five-point floating stars. 

In order to formulate (\ref{kA-system-eqn}) explicitly, we must first find a suitable approximation for the energy as a functional of density and orientation. 
Denote the density at a point $\bx$ as  $\rho(\bx)$ and orientation as $\sigma(\bx)$. If the energy between particles due to density is given by the (half)-sum of binary interactions (which is a suitable approximation for dilute states), the 'density' part of the energy gives 
\begin{equation} 
E_\rho=\frac{1}{2} \int \rho(\bx) \rho(\bx ') G(\bx-\bx') \mbox{d} \bx \mbox{d} \bx' 
\label{Erho} 
\end{equation} 
For a given $\sigma(\bx)$, the orientation represents a vector with cartesian coordinates 
$(\cos \sigma(\bx), \sin \sigma(\bx))$. If we assume that a function $\boldsymbol{H}\left( \sigma(\bx),\sigma(bx') \right)$ 
describes interaction of orientations, then the energy due to orientations is given by 
\begin{equation} 
E_\sigma=\frac{1}{2} \int \rho(\bx) \rho(\bx ') \boldsymbol{H}\left(\sigma(\bx),\sigma(\bx') \right) \mbox{d} \bx \mbox{d} \bx' 
\label{Esigma0} 
\end{equation} 
One possible approximation would be to assume that the interaction function $\boldsymbol{K}$ is simply proportional to the scalar product between $\sigma(\bx)$ and $\sigma(\bx')$. However, for the case of interacting stars or any objects with an $m$-fold symmetry, we must take into account that  $\sigma(\bx)$ and $\sigma(\bx')$ 
can differ by $2 \pi/m$ without affecting the interaction energy. Thus, the following form of the interaction  energy is proposed: 
\begin{equation} 
E_\sigma=\frac{1}{2}
 \int \rho(\bx) \rho(\bx ') K \big( |\bx-\bx' | \big) 
 \cos\left[ m \big(\sigma(\bx) - \sigma(\bx') \big)  \right]
  \mbox{d} \bx \mbox{d} \bx' 
\label{Esigma} 
\end{equation} 
Here, $K(r)$ is a scalar smoothing function, which must be determined from physical reasoning. An interesting case relevant for the evolution of the `pointy' objects is that the orientation part of the energy $E_\sigma$ is of much longer range than the density part $E_\rho$. 
The total energy is then simply
\begin{equation} 
E[\kappa]=E_\rho+E_\sigma,
\end{equation}
where $\kappa=(\rho,\sigma)$. 
\\

In Figure~\ref{fig:gyrons} we present a numerical simulation for the evolution in two dimensions of a collection of  five-point stars ($m=5$). In this case, there are two unknown variables: the mass density $\rho$, which multiplies the basis vector of $e_0$, and orientation $\sigma$ which is the coefficient of the basis vector $e_1$. The initial conditions for the evolution are taken to be a gaussian 
distribution of initial density, possessing random initial orientations (Fig.~\ref{fig:gyrons}, top left). In the simulation, we observe the formation of large-scale structures which are predominantly either individual  line segments or arrangements of three line segments meeting at a point to form a star-like structure (Fig.~\ref{fig:gyrons}, top right). An experiment, conducted by P.~D.~Weidman, started with 4mm plastic floating stars distributed within a circular shape with random initial orientation (Fig.~\ref{fig:gyrons}, bottom left). The stars are positioned on a net which is lowered slowly into the fluid. Care is taken to avoid residual water motion and convection. After 1 hour, the stars were found to assemble into line shapes or several-armed structures. Three-armed structures seem to be most common  (Fig.~\ref{fig:gyrons}, bottom right). While the position of the large-scale formations in the experiment is random and therefore impossible to predict analytically, the predominant shapes found in the experiment are reproduced in simulations using our model. 
\remfigure{ 
\begin{figure} [h]

\centering 
\includegraphics[width=16cm]{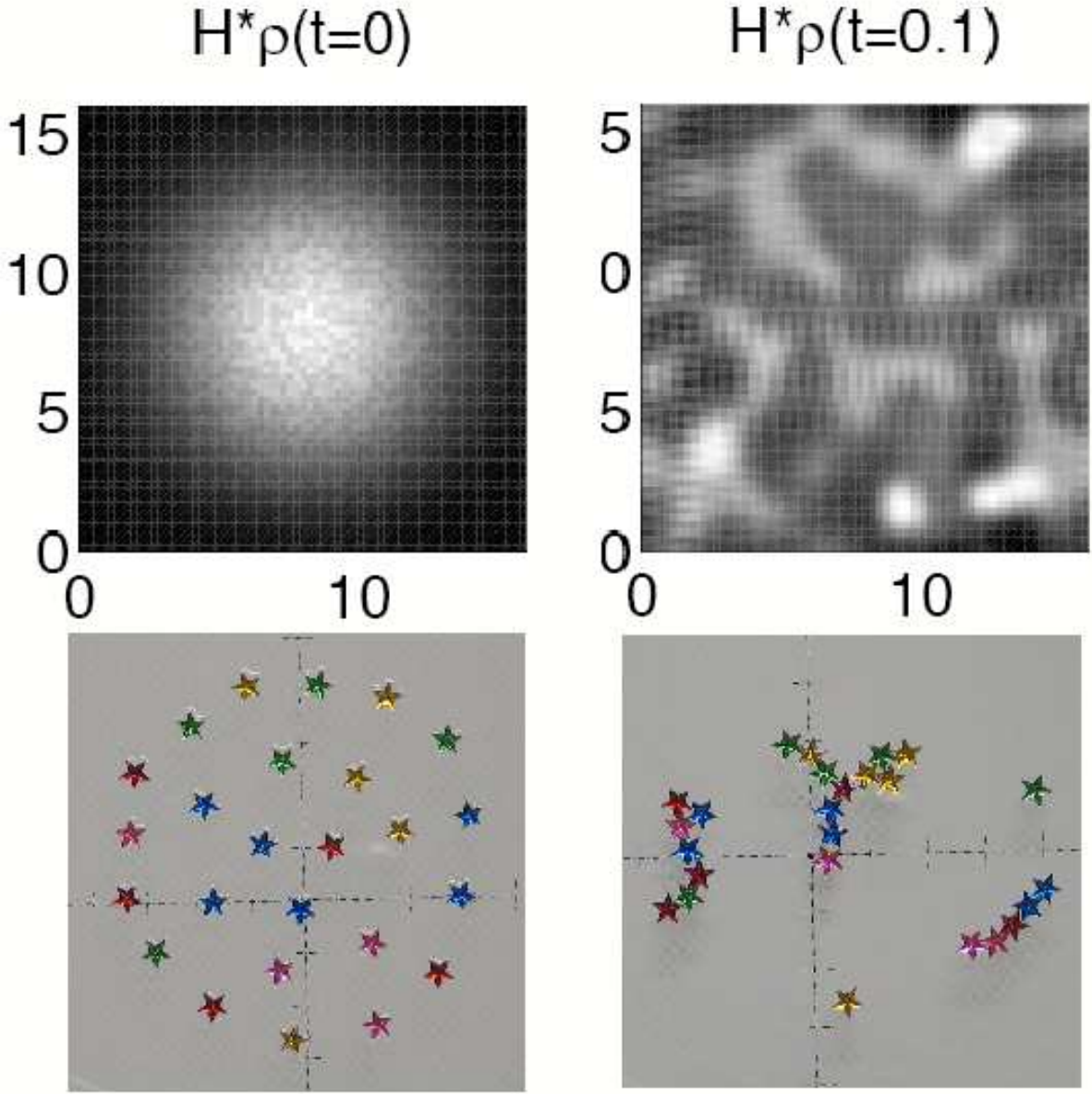} 
 \caption{ 
 Top:  Numerical simulation of density $\overline{\rho}=H*\rho$ where 
 $H(\mathbf{r})=e^{-\mathbf{r}^2}$ is a Gaussian filter. White denotes the areas with high density whereas black marks the areas with small density. Left: initial state for density, taken to be Gaussian:$\rho(\mathbf{r},0)= \rho_{0,k} e^{-|\mathbf{r}|^2/l_\rho^2}$. In the simulation, the particles are assumed to have a five-fold symmetry. The initial conditions for orientation are random. Top right: final state.  
 Bottom: an experiment reproducing this numerical solution 
 (courtesy of P.D.Weidman). 
Bottom left: initial conditions - five-point stars are positioned on a net and slowly lowered into the fluid. Initial orientations for the stars are random. Bottom right: final state after 1 hr. 
  }
   \label{fig:gyrons}
 \end{figure} 
 } 

\section{Acknowledgements} 
We are indebted to Cesare Tronci for his insightful comment about a key step in the derivation of the GOP equation, that made the diamond-form of Darcy's law more generally applicable. This improvement will be pursued elsewhere.
We are grateful for stimulating discussions with P. D. Weidman, whose remarkable experiments on anisotropic floating particles set this work in motion. We are also grateful for the permission to use  Prof. Weidman's experimental results in this work.
We are also grateful to S. R. J. Brueck, who introduced us to the fascinating world of nanoscale self assembly. The authors were partially supported by NSF grant NSF-DMS-05377891. 
The work of DDH was partially supported  by the Royal Society of London and the US Department 
of Energy Office of Science Applied Mathematical Research. VP acknowledges the support of von Humboldt foundation and the hospitality of the Institute for Theoretical Physics, University of K\"oln where this work was completed. 


\end{document}